\documentclass[review,onefignum,onetabnum]{siamart171218}

\usepackage{braket,amsfonts}
\usepackage{array}
\usepackage[caption=false]{subfig}
\usepackage{pgfplots,bm,multirow}

\newsiamthm{claim}{Claim}
\newsiamremark{remark}{Remark}
\newsiamremark{hypothesis}{Hypothesis}
\crefname{hypothesis}{Hypothesis}{Hypotheses}

\usepackage{algorithmic}

\usepackage{graphicx,epstopdf}

\Crefname{ALC@unique}{Line}{Lines}

\usepackage{amsopn}

\usepackage{xspace}
\usepackage{bold-extra}
\usepackage[most]{tcolorbox}

\colorlet{texcscolor}{blue!50!black}
\colorlet{texemcolor}{red!70!black}
\colorlet{texpreamble}{red!70!black}
\colorlet{codebackground}{black!25!white!25}

\lstdefinestyle{siamlatex}{%
  style=tcblatex,
  texcsstyle=*\color{texcscolor},
  texcsstyle=[2]\color{texemcolor},
  keywordstyle=[2]\color{texemcolor},
  moretexcs={cref,Cref,maketitle,mathcal,text,headers,email,url},
}

\tcbset{%
  colframe=black!75!white!75,
  coltitle=white,
  colback=codebackground, 
  colbacklower=white, 
  fonttitle=\bfseries,
  arc=0pt,outer arc=0pt,
  top=1pt,bottom=1pt,left=1mm,right=1mm,middle=1mm,boxsep=1mm,
  leftrule=0.3mm,rightrule=0.3mm,toprule=0.3mm,bottomrule=0.3mm,
  listing options={style=siamlatex}
}

\newtcblisting[use counter=example]{example}[2][]{%
  title={Example~\thetcbcounter: #2},#1}

\newtcbinputlisting[use counter=example]{\examplefile}[3][]{%
  title={Example~\thetcbcounter: #2},listing file={#3},#1}

\DeclareTotalTCBox{\code}{ v O{} }
{ 
  fontupper=\ttfamily\color{black},
  nobeforeafter,
  tcbox raise base,
  colback=codebackground,colframe=white,
  top=0pt,bottom=0pt,left=0mm,right=0mm,
  leftrule=0pt,rightrule=0pt,toprule=0mm,bottomrule=0mm,
  boxsep=0.5mm,
  #2}{#1}

\patchcmd\newpage{\vfil}{}{}{}
\begin{tcbverbatimwrite}{tmp_\jobname_header.tex}
\title{Fast-Converging and Asymptotic-Preserving DSMC 
\thanks{Submitted to the editors DATE.
}}

\author{Bin Hu, 
    Liyan Luo,  
    Kaiyuan Wang, 
	Lei Wu\thanks{Department of Mechanics and Aerospace Engineering, Southern University of Science and Technology, Shenzhen 518055, China (\email{wul@sustech.edu.cn}) }
}

\headers{Direct Intermittent GSIS}{Hu, Luo, Wang, Wu}
\end{tcbverbatimwrite}
\input{tmp_\jobname_header.tex}

\ifpdf
\hypersetup{pdftitle={Guide to Using  SIAM'S \LaTeX\ Style} }
\fi

\definecolor{mygreen}{rgb}{0.0,0.55,0.55}

\nolinenumbers

\begin{document}
\maketitle

\begin{tcbverbatimwrite}{tmp_\jobname_abstract.tex}
\begin{abstract}
Improving the efficiency of the direct simulation Monte Carlo (DSMC) method has become increasingly urgent with the rapid development of space exploration. To address this issue, the direct intermittent general synthetic iteration scheme (DIG) has recently been proposed to enable DSMC's rapid and accurate convergence to steady-state solutions, even when the cell size is much larger than the mean free path in near-continuum flow regimes. The first part of the paper is devoted to the mathematical analysis of DIG's fast-converging and asymptotic-preserving properties. Because the Boltzmann equation is analytically intractable, the analysis is conducted using the linearized Bhatnagar–Gross–Krook model. It is found that, in the near-continuum flow regime, the DIG method asymptotically recovers the Navier–Stokes equations when the cell size is $\Delta x\sim \mathcal{O}(1)$, rather than being constrained by the mean free path. Moreover, after a single cycle of DIG evolution, the deviation from the final steady-state solution is reduced at least by a factor of five.
In the second part of the paper, the Poiseuille flow and hypersonic flow passing over cylinder are investigated using the DIG scheme, with different time step and cell sizes, thereby demonstrating its efficiency and accuracy in  multiscale flow simulation. Specifically, when the Knudsen number is 0.01, the DIG method is found to be faster than the traditional DSMC method by two orders of magnitude. The performance gain becomes even greater at smaller Knudsen numbers. The proposed method holds great potential for engineering applications.
\end{abstract}


\end{tcbverbatimwrite}
\input{tmp_\jobname_abstract.tex}

\section{Introduction} \label{sec:introduction}

Rarefied gas flows play a crucial role in modern engineering applications such as spacecraft re-entry, micro-electromechanical systems, controlled nuclear fusion, and vacuum technology \cite{chen-1984,Karniadakis2005,Wu2022Book}. The degree of rarefaction effects is characterized by the dimensionless Knudsen number, defined as the ratio of the mean free path $\lambda$ of the gas molecules to the characteristic flow length $L$:
\begin{equation}\label{Kn_original}
	\text{Kn}=\frac{\lambda}{L}, 
    \quad\text{with}\quad
	\lambda=\frac{\mu}{p_0}\sqrt{\frac{\pi{}k_BT_0}{2m_g}},
\end{equation}
where $\mu$ is the shear viscosity, $p_0$ is the gas pressure, $k_B$ is the Boltzmann constant, $T_0$ is the gas temperature, and $m_g$ is the molecular mass. 
The Navier–Stokes (NS) equations fail to accurately describe such flows because they assume a continuous medium and local thermodynamic equilibrium. The Boltzmann equation, which accounts for molecular-level interactions and non-equilibrium effects, must be used instead. 


Numerical simulation of the Boltzmann equation presents a significant challenge. For a monatomic gas, the velocity distribution function is defined in a six-dimensional phase space, and the binary collision operator involves a fivefold nonlinear integral. The cases of polyatomic gases and chemical reaction are more complex, as additional internal degrees of freedom must be taken into account \cite{WangCS}. The direct simulation Monte Carlo (DSMC) method \cite{bird-1994} has been widely used as a stochastic approach that reproduces the streaming and collision processes in the Boltzmann equation \cite{wagner1992convergence}. Its popularity stems from the use of a significantly smaller number of simulation particles compared with the enormous number of velocity nodes required in the discrete velocity method \cite{broadwell1964,chu1965kinetic,sone1981discrete,aristov2001direct}, as well as from its flexibility in incorporating complex physical and chemical processes.
The method is highly efficient in simulating rarefied flows with $\text{Kn}>0.1$, such as those encountered in high-altitude or hypersonic aerodynamics, where molecular collisions are infrequent. However, in the near-continuum regime ($\text{Kn}<0.01$), where intermolecular collisions dominate, DSMC becomes computationally expensive. In this case, the time step and cell size must be smaller than the mean collision time and mean free path, respectively, resulting in prohibitively high computational costs. 
Moreover, in low-speed flow simulations, the intrinsic statistical noise of the DSMC requires a large amount of averaging to capture the subtle flow signals \cite{Hadjiconstantinou2003JCP}.

To remove the deficiency of DSMC in the near-continuum regime, the NS–DSMC coupling method is proposed first, where the NS equations and DSMC are applied in the continuum and rarefied regions, respectively~\cite{schwartzentruber-2006,schwartzentruber-2007}. Information between the two regions is exchanged through a buffer zone, where macroscopic flow quantities from NS are used to generate particle boundary conditions for DSMC, and microscopic fluxes from DSMC are averaged and fed back to the NS solver. Despite their success, two major drawbacks persist: the interface between the rarefied and continuum regions is difficult to determine a priori, and DSMC converges slowly near the interface when the Knudsen number is small. 
Although, in theory, the second problem can be alleviated by replacing the NS equations with high-order moment equations~\cite{yang2020hybrid}, their application to hypersonic flows remains limited, since the moment system \cite{Grad1949} can produce nonphysical solutions \cite{Struchtrup2005}.

To avoid domain decomposition, asymptotic-preserving (AP) schemes have been developed (see the relevant review papers~\cite{ADAMS20023,Jin2022_APschemes}), allowing the use of large time steps and cell sizes while preserving accuracy across a wide range of gas rarefaction. In the DSMC framework, AP schemes include the moment-guided~\cite{Degond2011}, time-relaxed~\cite{pareschi-2001}, exponential Runge–Kutta~\cite{dimarco-2011}, and successive penalty time-discretization Monte Carlo methods~\cite{ren-2014}.
The key idea of these AP-DSMC methods is to decompose the collision operator into a stiff linear component and a less stiff nonlinear component. The stiff component is typically approximated using a relaxation-time model, which ensures the correct asymptotic behavior and consistency with the Euler equations in the continuum limit. However, these methods often fail to accurately recover transport coefficients~\cite{JIN199551}.
Recently, an AP-DSMC has been developed by Fei~\cite{fei-2023,FEI2025114196}, which achieves both asymptotic preservation of the NS equations and second-order accuracy in the fluid regime. The key idea is that the artificial numerical dissipation introduced by large cell sizes \cite{Garcia1998} and time steps \cite{Garcia2000} can be quantified exactly and then removed by appropriately modifying the collision operator \cite{Fei2020Zhang}. At large Knudsen numbers, however, this modification must be switched off.

Despite these advances, DSMC’s full potential remains under-utilized. First, AP schemes should be simple and easily extendable to applications such as hypersonic chemically reacting flows. 
Second, for most steady-state problems in rarefied gas dynamics, tracking time evolution is unnecessary. Beyond AP-DSMC schemes that allow large time step and cell size, methods should accelerate convergence to the steady state, avoiding unnecessary temporal evolution.

Within the discrete velocity framework, AP schemes, such as the unified gas-kinetic schemes~\cite{xu-2010,guo-2013,YuanLiuZhong2021} and general synthetic iterative scheme (GSIS)~\cite{CWF,Zhang2023_GSIS}, where the kinetic equations are solved together with the macroscopic synthetic equations throughout the entire computational domain, have demonstrated engineering applications. Specifically, in GSIS, solving the macroscopic synthetic equations directly to steady-state solution eliminates the need for many intermediate evolution steps.
Very recently, the core concept of GSIS has been extended to DSMC, leading to the development of a direct intermittent GSIS scheme known as DIG \cite{DIG}. In DIG, the traditional DSMC simulation is intermittently enhanced by a macroscopic synthetic equation, whose constitutive relations for stress and heat flux are extracted from the DSMC. These equations are solved to obtain the steady-state solution, which is then used to update the particle distribution in DSMC. This two-way coupling not only accelerates convergence to the steady state via the enhanced global flow information exchange\footnote{Recent work shows that applying a similar GSIS idea locally can eliminate the need to solve macroscopic equations, thereby accelerating convergence. However, an additional multigrid method is employed to achieve maximum speed-up \cite{cai_hu2025}.} but also asymptotically preserves the NS limit in the continuum flow regime, allowing the cell size to be much larger than the molecular mean free path. Importantly, the stochastic flexibility of the original DSMC method is preserved. Moreover, DIG has been easily extended to gas-mixture flows with disparate molecular masses \cite{Luo2025_APDSMCMixtures} and holds strong potential for direct application to flow simulations involving complex physical and chemical processes.

This paper is devoted to a mathematical analysis of DIG's fast-converging and AP properties. To enable analytical investigation, we employ the Bhatnagar–Gross–Krook (BGK) kinetic equation, which mimics the behavior of DSMC while retaining the essential features of rarefied gas dynamics. The corresponding behavior of the DSMC method accelerated by DIG is subsequently illustrated through numerical simulations.

The remainder of this paper is organized as follows. In Section~\ref{Kinetic_model}, we introduce the linearized BGK model \cite{BGK1954} and present the Fourier analysis of the conventional iterative scheme,  GSIS, and DIG. In Section~\ref{sec:ap}, the AP property is analyzed based on the Chapman-Enskog expansion. Section~\ref{sec:num} presents numerical simulations of the Poiseuille flow and hypersonic flow over cylinder to demonstrate the fast-converging and AP properties of DIG, which reconcile with the analytical results based on the BGK model. Finally, Section~\ref{sec:conclusion} concludes the paper.

\section{The linearized BGK equation and convergence analysis} \label{Kinetic_model}

If the gas deviates slightly from the global equilibrium, the velocity distribution function $f(t,\bm{x},\bm{v})$ can be linearized around global equilibrium distribution 
\begin{equation}
    f_{eq} = \frac{{\exp}(-v^2)}{\pi^{3/2}} 
\end{equation}
as $f = f_{eq}+ h$, where $h(t,\bm{x},\bm{v})$ is the perturbation velocity distribution function satisfying the linearized BGK model
\begin{equation}\label{linearized_equation}
    \begin{aligned}
        \frac{\partial h}{\partial t}+\bm v \cdot \frac{\partial h}{\partial \bm x}= \underbrace{\delta_{r p}\left[
         \rho+2 \bm u \cdot \bm v+\tau\left(v^2-\frac{3}{2}\right)
        \right] f_{eq}-\delta_{r p} h }_{\mathcal{L}(h)},
    \end{aligned}
\end{equation}
with $t$ the time normalized by $L/v_m$, $\bm{x}=(x_1,x_2,x_3)$ the three-dimensional Cartesian coordinates normalized by $L$, and $\bm{v}=(v_1,v_2,v_3)$ the three-dimensional molecular velocity normalized by $v_m$, where $v_m=\sqrt{2k_BT_0/m_g}$ is the most probable speed. The rarefaction parameter $\delta_{rp}$ is related to the Knudsen number as $\sqrt{\pi}/2\text{Kn}$, and can be viewed as the mean collision frequency.

The macroscopic quantities, such as the perturbation density $\rho$ normalized by $p_0m_g/k_BT_0$, flow velocity $\bm u=(u_1,u_2,u_3)$ normalized by $v_m$, perturbation temperature $\tau$ normalized by $T_0$, stress tensor $\bm{\sigma}$ normalized by $p_0$, and heat flux $\bm{q}=(q_1,q_2,q_3)$ normalized by $p_0v_m$, are defined as the velocity moments of $h$:
\begin{equation}\label{macro}
    \begin{aligned}
        &[\rho, \bm{u}, \tau] =\iiint \phi(v) h d \bm v, ~\text{with}~ \phi(\bm v)=\left[1, \bm{v}, \frac{2}{3} v^2-1\right],\\
    & \sigma_{l j}=2 \iiint v_{\langle l} v_{j \rangle} h d \bm v \equiv 2 \iiint\left(v_l v_j-\frac{v^2}{3} \delta_{l j}\right) h d \bm v, \\
  & \bm q=\iiint\left(v^2-\frac{5}{2}\right) \bm v h d \bm v,
    \end{aligned}
\end{equation}
where the integration is carried out over the entire velocity space $\mathbb{R}^3$, $\langle\cdot\rangle$ represent the traceless tensor, and $\delta_{lj}$ is the Kronecker delta function, with $l,j=1,2,3$.

\subsection{Conventional iterative scheme}

The original DSMC method is first-order accurate in time, since the streaming and collision are treated sequentially. The same approach can be applied to the linearized BGK model~\eqref{linearized_equation}:
\begin{equation}\label{Lie_splitting0}
    \left\{
\begin{aligned}
\frac{\partial h}{\partial t} =\mathcal{L}(h),
&\quad\Rightarrow \frac{h^\ast-h^k}{\Delta t}=\mathcal{L}(h^k) +\mathcal{O}(\Delta t)\frac{\partial^2h}{\partial t^2},\\
\frac{\partial h}{\partial t} + \bm{v} \cdot \frac{\partial h}{\partial\bm{x}}  = 0,
&\quad\Rightarrow  h^{k+1}(\bm{x}+\bm{v} \Delta t, \bm{v}) = h^{\ast}(\bm{x}, \bm{v}).
\end{aligned}
\right.
\end{equation}
Note that each equation is approximated using a numerical scheme with first-order global temporal accuracy. For the collision operator, given the velocity distribution function $h^k$ at $t=t_k$, the forward Euler scheme is employed to compute the intermediate distribution function $h^\ast$. For the streaming term, the analytical solution for the velocity distribution function $h^{k+1}$ at $t=t_k+\Delta t$ is first derived from $h^\ast$ and then expanded in a Taylor series with first-order global temporal accuracy, resulting in
\begin{equation}\label{Lie_splitting}
  h^\ast= h^{k+1}(\bm{x},\bm{v})+{\Delta t} \bm{v} \cdot \frac{\partial h^{k+1}(\bm{x},\bm{v})}{\partial\bm{x}} +\mathcal{O}(\Delta^2 t)\frac{\partial^2 h(\bm{x},\bm{v})}{\partial x_l \partial x_j}v_lv_j ,
\end{equation}
where the Einstein summation is used with the dummy indexes $l$ and $j$.
Thus, the first-order temporal discretization of \eqref{linearized_equation}, analogous to that of the DSMC method, can be expressed as:
\begin{equation}
\label{linear_eq_discretization_t}
    \frac{h^{k+1}(\bm{x}, \bm{v})-h^k(\bm{x}, \bm{v})}{\Delta t} +{\bm v} \cdot\frac{\partial h^{k+1}(\bm{x}, \bm{v})}{\partial \bm x}
    =\mathcal{L}(h^{k}(\bm{x}, \bm{v})).
\end{equation}
For simplicity, the spatial derivative is left in its continuous form; its influence on the asymptotic-preserving property will be analyzed in Section~\ref{sec:ap}. 

It is interesting to note that, when $\delta_{rp} \Delta t =1$, \eqref{linear_eq_discretization_t} becomes 
   $ \delta_{rp}h^{k+1}+{\bm v} \cdot\frac{\partial h^{k+1}}{\partial \bm x}
    =\delta_{rp}h^{k}+\mathcal{L}(h^k)$,
which is known as the conventional iterative scheme (CIS). For this reason, the scheme in \eqref{linear_eq_discretization_t} is also referred to as the CIS.
When the entire process reaches a steady state $h_s(\bm{x},\bm{v})$, the governing equation reduces to
\begin{equation}\label{steady_state}
    \begin{aligned}
        {\bm v} \cdot\frac{\partial h_s}{\partial \bm x}=\mathcal{L}(h_s).
    \end{aligned}
\end{equation}

We are interested in examining how rapidly the error $Y^{k+1}(\bm x, \bm v)=h^{k+1}(\bm x, \bm v)-h_s(\bm x, \bm v)$ decays as  the iteration index $k$ increases.
Correspondingly, errors associated with the macroscopic quantities $M=[\rho, \bm u, \tau]$ are defined as
\begin{equation}\label{phi_k_1}
   \begin{aligned}
    \Phi^{k+1}(\bm x) 
    =M^{k+1}(\bm x)-M_s(\bm x)=\iiint Y^{k+1}(\bm x, \bm v) \phi(\bm v) d \bm v.
\end{aligned} 
\end{equation}
In order to determine the error decay rate $e_c$, we perform the Fourier stability analysis by seeking the eigenfunctions $\bar{Y}(v)$ and $\alpha_M=\left[\alpha_{\rho}, \alpha_{\bm u}, \alpha_\tau\right]$ of the following forms:
\begin{equation}
    \begin{aligned}
    Y^{k+1}(\bm{x}, \bm v)&= e^k_c \bar{Y}(e_c,\bm v) \exp (i \bm{\theta} \cdot \bm{x}), \\
     \Phi^{k+1}_M(\bm{x})&=  e^{k}_c \alpha_{M,c} \exp (i \bm{\theta} \cdot \bm{x}),
\end{aligned}\label{Y_and_phi}
\end{equation}
where $\bm{\theta}=\left(\theta_1, \theta_2, \theta_3\right)$ is the wave vector of perturbation satisfying $|\bm\theta|=1$, and $i$ is the imaginary unit. The numerical scheme \eqref{linear_eq_discretization_t} is unstable when the error decay rate is larger than unity, while slow (fast) convergence occurs when $|e_c|$ approaches one (zero). 

It can be derived from  \eqref{linear_eq_discretization_t} to \eqref{Y_and_phi} that
$\alpha_M=\iiint \bar{Y}(e_c,\bm v) \phi(\bm v) d \bm v$, where
\begin{equation}\label{barY_v}
\begin{aligned}
   \bar{Y}(e_c,\bm {v}) &= \left[
        \alpha_{\rho} + 
        2 {\alpha}_u \cdot \bm{v} + 
        \alpha_\tau \left(v^2 - \frac{3}{2}\right) 
    \right] y_0(e_c,\bm{v}), \\
    \text{with}\quad
     y_0(e_c,\bm v)&=\frac{f_{{eq}}}{1+(e_c-1)\left( \Delta t \delta_{rp}\right)^{-1}+ie_c \delta_{r p}^{-1} \bm{\theta} \cdot \bm{v}}.
\end{aligned}
\end{equation}
On multiplying~\eqref{barY_v} with $\phi(v)$ defined in \eqref{macro} and integrating the resultant equations with respect to the molecular velocity $\bm v$, we obtain five linear algebraic equations for five unknown elements in $\alpha_M$. These algebraic equations can be written in the matrix form as 
\begin{equation}\label{CIS_error_decay_matrix}
C_5 \alpha_{M,c}^{\top}= \alpha_{M,c}^{\top}, \quad\text{with}~
        C_5=\iiint\left[1,2 \bm v, v^2-\frac{3}{2} \right]^{\top} \phi(\bm v) y_0(e_c,\bm v) d \bm v,
\end{equation}
where the superscript $\top$ is the transpose operator, and $C_5$ is $5 \times 5$ matrix that contains the eigenvalue $e_c$.

\begin{figure}
    \centering
    \includegraphics[width=0.7\linewidth]{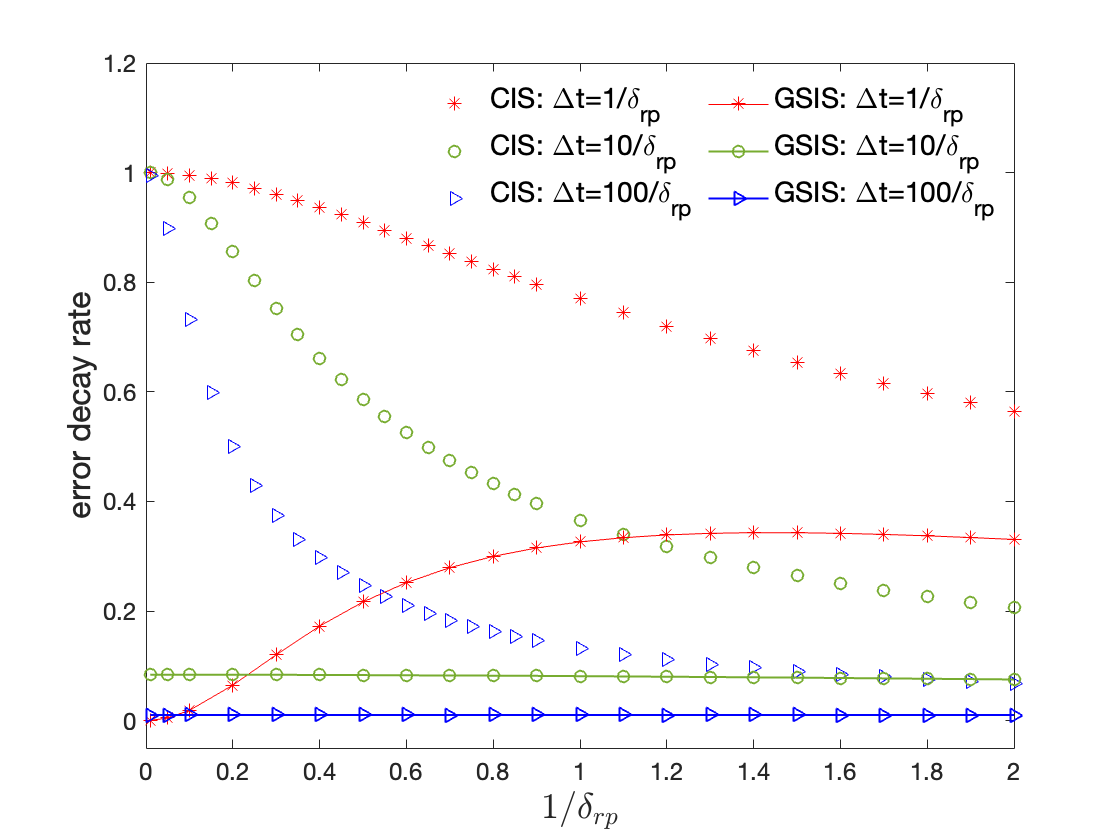}
    \caption{Error decay rates as functions of the Knudsen number and time step for CIS and GSIS.}
    \label{fig:CIS_GSIS}
\end{figure}

The error decay rate can be obtained by numerically solving $\det(C_5-I_{5\times5})=0$, where $I_{5\times5}$ is a ${5\times5}$ identity matrix.
Specifically, when $\delta_{rp} \Delta t =1$, $e_c$ corresponds to the eigenvalue of the matrix $e_cC_5$, which can be readily computed using standard eigenvalue algorithms. Otherwise, the Newton–Raphson method is employed to determine $e_c$.
Results of error decay rate as functions of the Knudsen number and time step are shown in Fig. \ref{fig:CIS_GSIS}. It is clear that when the Knudsen number is large, $e_c$ goes to zero so that the error decays quickly. This means that CIS is rather efficient for highly rarefied gas flows, i.e., the converged solution can be found within dozens of iterations. On the contrary, when $\text{Kn} \rightarrow 0$, $|e_c| \rightarrow 1-\mathcal{O}(\text{Kn}^2)$, meaning that the CIS is extremely slow in the (near) continuum flow regime. 
Also note that $|e_c|$ decreases when the time step increases. 

\subsection{The GSIS scheme}

In the GSIS framework, macroscopic synthetic equations are constructed to guide the system rapidly toward the steady state, bypassing unnecessary intermediate stages. Among the two versions of GSIS \cite{FCA,ZhuL-2021,zeng-2023-cicp}, the second version—which is better suited for DSMC—is employed.

First, given the velocity distribution function $h^k$ at the $k$-th time step, its value at the intermediate ($k+1/2$)-th step is obtained in a similar way as \eqref{linear_eq_discretization_t}:
\begin{equation}\label{GSIS_iteration}
    \frac{h^{k+1/2}-h^k}{\Delta t} +{\bm v} \cdot\frac{\partial h^{k+1/2}}{\partial \bm x}
       =\mathcal{L}(h^k).
\end{equation}

Second, based on $h^{k+1/2}$ and the corresponding macroscopic quantities $M^{k+1/2}$, we develop a procedure to determine the macroscopic quantities that are close to the steady state. To this end, we multiply the steady-state BGK equation \eqref{steady_state} by $1,2\bm v$, and $v^2-\frac{3}{2}$, respectively, and integrate the resultant equations with respect to the molecular velocity $\bm v$, obtaining:
\begin{equation}\label{GSIS_II}
  \left\{  \begin{aligned}
        & \frac{\partial u^{k+1}_l}{\partial x_l}=0, \\
&\frac{\partial \rho^{k+1}}{\partial x_j}+\frac{\partial \tau^{k+1}}{\partial x_j}+\frac{\partial \sigma^{k+1}_{l j}}{\partial x_l}=0, \quad j=1,2,3,\\
& \frac{\partial q^{k+1}_l}{\partial x_l}=0.
    \end{aligned}
    \right.
\end{equation} 
These macroscopic equations are not closed since the expressions for stress $\bm \sigma_{l j}$ and heat flux $\bm {q}$ are not known. 
In the second version of GSIS, the constitutive relations are decomposed into the continuum part (i.e., the Newton law of stress and the Fourier law of heat conduction) and the  non-continuum part (i.e., the bracketed terms represent high-order terms (HoTs) that account for rarefaction effects):
\begin{equation}\label{GSIS_II_HoTs}
\left\{
\begin{aligned}
    & \sigma_{l j}^{k+1}=-2 \delta_{rp}^{-1} \frac{\partial u_{<l}^{k+1}}{\partial x_{j>}}+\left(\sigma_{l j}^{k+1 / 2}+2 \delta_{rp}^{-1} \frac{\partial u_{<l}^{k+1 / 2}}{\partial x_{j>}}\right), \\
    & q_l^{k+1}=-\frac{5}{4 } \delta_{rp}^{-1} \frac{\partial \tau^{k+1}}{\partial x_l}+\left(q_l^{k+1 / 2}+\frac{5}{4 } \delta_{rp}^{-1} \frac{\partial \tau^{k+1 / 2}}{\partial x_l}\right).
\end{aligned}
\right.
\end{equation}
When the Knudsen number is small, HoTs are proportional to $\text{Kn}^2$ if the error of spatial discretizaiton is not account for. Consequently, the constitutive relations for the stress and heat flux reduce to the Newtonian stress law and Fourier’s law of heat conduction, respectively. When Kn is large, HoTs are important, which cannot be expressed analytically even with the high-order Chapman-Enskog expansion \cite{CE1970} or the Grad moment method \cite{Grad1949}; they can only be extracted by numerically solving the kinetic equation. 


To calculate the error decay rate of GSIS, the error functions are defined as
\begin{equation}\label{Y_and_phi_GSIS}
    \begin{aligned}
    & Y^{k+1/2}(\bm{x}, \bm v)=h^{k+1/2}(\bm x, \bm v)-h_s(\bm x, \bm v)=e^k_g \bar{Y}(e_g,\bm v) \exp (i \bm{\theta} \cdot \bm{x}), \\
    & \Phi^{k+1}_M(\bm{x})=M^{k+1}(\bm x)-M_s(\bm x)=e^{k}_g \alpha_{M,g} \exp (i \bm{\theta} \cdot \bm{x}).
\end{aligned}
\end{equation}
Note that, unlike the CIS, the macroscopic quantities $M^{k+1}$ are calculated from the synthetic equations \eqref{GSIS_II}, rather than directly from the distribution function $h^{k+1/2}$. 

After some algebraic manipulation, we find that the form of $\bar{Y}$ remains unchanged. 
The errors,  $\Phi$ in \eqref{phi_k_1}, are governed by \eqref{GSIS_II_HoTs} and \eqref{Y_and_phi_GSIS}, with the macroscopic quantities replaced by their corresponding error functions.
Therefore, the error decay rate can be obtained by solving the following linear systems:
\begin{equation}\label{GSIS_matrix}
\left\{
    \begin{aligned}
        & i \theta_l \alpha_{u,l}=0, \\
            & i \theta_j\left(\alpha_{\rho}+\alpha_\tau\right)
            +\frac{1}{\delta_{rp}}\alpha_{u,j}+\frac{\theta_j}{3 \delta_{rp}} \theta_l \alpha_{u,l}=S_{u,j}, \quad j=1,2,3,\\
        &\frac{5}{4\delta_{rp}}\alpha_\tau =S_{\tau},
    \end{aligned}
    \right.
\end{equation}
where the source terms, originating from the HoTs and evaluated at the intermediate iteration step $k+1/2$, are
\begin{equation}\label{source_term_GSIS}
    \begin{aligned}
        & S_{u,j}=\iiint\left[\frac{v_j}{\delta_{rp}}+\frac{\theta_j}{3 \delta_{rp}} \theta_l v_l-2 i \theta_l v_{\langle j} v_{l\rangle}\right] \bar{Y}(e_g,\bm v) d \bm v, \\
        & S_{\tau}=\iiint\left[\frac{5}{4 \delta_{rp} } \left(\frac{2}{3} v^2-1\right)
        -i\theta_lv_l\left(v^2-\frac{5}{2}\right)\right]\bar{Y}(e_g,\bm v) d \bm v.
    \end{aligned}
\end{equation}

\eqref{GSIS_matrix} and \eqref{source_term_GSIS} can be rearranged into the matrix form as $L_5\alpha_{M,g}^{\top} = R_5(e_g)\alpha_{M,g}^{\top}$, and the error decay rate can be obtained by numerically computing $\det(L_5^{-1}R_5-I_{5\times5})=0$. 
Results are shown in Fig.~\ref{fig:CIS_GSIS}. In the examined range of Kn, the GSIS converges faster than the CIS.  When $\Delta t = \delta^{-1}_{rp}$, the traditional GSIS is recovered \cite{FCA}, for which the error decay rate $|e_g|\rightarrow\mathcal{O}(\text{Kn}^2)$ when $\text{Kn}\rightarrow0$. 
As the time step increases, the error decay rate generally decreases for most Knudsen numbers, except in the vicinity of $\text{Kn} \sim 0$, where $|e_g|$ does not vanish but remains at a small value.
This does not raise any problems in the efficient simulation of near-continuum flows, since here the convergence analysis is performed in an infinite domain. In practice, however, wall confinement leads to the formation of a Knudsen layer, which results in an effective $\text{Kn} \sim 1$. This implies that only the maximum error decay rate is of practical significance \cite{CWF,Wu2022Book}. Thus, generally speaking, the convergence speed in both CIS and GSIS increases with time step. 

A subtle point to note is that when $\Delta t = \delta^{-1}_{rp}$, the error decay rate in GSIS does not vanish at large Knudsen numbers. In this case, a limiter can be applied when updating the macroscopic quantities $M$ at the $(k+1)$-th iteration step \cite{CWF}, e.g.
\begin{equation}\label{limit_macro}
        M^{k+1} = \beta M_{\text{macro}} + (1-\beta)M^{k+1/2},
        \quad\text{with}~  \beta = \frac{\text{min}(\text{Kn},\text{Kn}_{\text{th}})}{\text{Kn}},
\end{equation}
where $M_\text{macro}$ is the solution of \eqref{GSIS_II} and $\text{Kn}_{\text{th}}=1$.
For larger time steps, e.g. $\Delta t = 10/\delta_{rp}$ and $\Delta t = 100/\delta_{rp}$, the error decay rate goes to zero at large Knudsen numbers; however, it is still recommended to apply the limiter.

\subsection{The DIG scheme}\label{sec:DIG}

The DIG method was originally proposed to accelerate DSMC simulations \cite{DIG, Luo2025_APDSMCMixtures}. Because the stochastic nature of DSMC inherently introduces noise, a single DIG cycle consists of $m-1$ DSMC steps, during which the constitutive relations are obtained by averaging to reduce the stochastic noise, followed by one GSIS step. To analyze its convergence, we construct an analogous DIG procedure based on the linearized BGK equation. Specifically, the CIS is applied for $m-1$ consecutive steps, followed by one GSIS step, using the following constitutive relation:
\begin{equation}\label{DIG_HoTs}
\left\{
\begin{aligned}
    & \sigma_{l j}^{m+1}=-2 \delta_{rp}^{-1} \frac{\partial u_{<l}^{m+1}}{\partial x_{j>}}+\left(\bar{\sigma}_{l j}+2 \delta_{rp}^{-1} \frac{\partial \bar{u}_{<l}}{\partial x_{j>}}\right), \\
    & q_l^{m+1}=-\frac{5}{4 } \delta_{rp}^{-1} \frac{\partial \tau^{m+1}}{\partial x_l}+\left(\bar{q}_l+\frac{5}{4 } \delta_{rp}^{-1} \frac{\partial \bar{\tau}}{\partial x_l}\right),
\end{aligned}
\right.
\end{equation} 
where the overbar in brackets denotes some kind of average (usually the arithmetic mean) over the previous $m-1$ CIS steps, which mimics the suppression of statistical fluctuations in DSMC.

Similar to the CIS and GSIS, the convergence rate of the DIG for the linearized BGK equation is evaluated by defining the following error functions: 
\begin{equation}
\begin{aligned}
 & \left\{
\begin{aligned}
     Y^{2}(\bm{x}, \bm v)&=h^{2}(\bm x, \bm v)-h_s(\bm x, \bm v)=e_{c} \bar{Y}(e_c,\bm v) \exp (i \bm{\theta} \cdot \bm{x}),  \\
    \Phi^{2}_M(\bm{x})&=M^{2}(\bm x)-M_s(\bm x)=e_{c} \alpha_{M,c} \exp (i \bm{\theta} \cdot \bm{x}),\\
     Y^{3}(\bm{x}, \bm v)&=h^{3}(\bm x, \bm v)-h_s(\bm x, \bm v)=e_{c}^2 \bar{Y}(e_c,\bm v) \exp (i \bm{\theta} \cdot \bm{x}), \\ \Phi^{3}_M(\bm{x})&=M^{3}(\bm x)-M_s(\bm x)=e_{c}^{2} \alpha_{M,c} \exp (i \bm{\theta} \cdot \bm{x}),\\
    & ~\vdots \\
    Y^{m}(\bm{x}, \bm v)&=h^{m}(\bm x, \bm v)-h_s(\bm x, \bm v)=e_{c}^{m-1} \bar{Y}(e_c,\bm v) \exp (i \bm{\theta} \cdot \bm{x}), \\
    \Phi^{m}_M(\bm{x})&=M^{m}(\bm x)-M_s(\bm x)=e_{c}^{m-1} \alpha_{M,c} \exp (i \bm{\theta} \cdot \bm{x}),
\end{aligned}
    \right.
    \\
 & \left\{
\begin{aligned}
    Y^{m+1/2}(\bm{x}, \bm v)&=h^{m+1/2}(\bm x, \bm v)-h_s(\bm x, \bm v)=e_{d}e_{c}^{m-1} \bar{Y}(e_d,\bm v) \exp (i \bm{\theta} \cdot \bm{x}), \\
    \Phi^{m+1}_M(\bm{x})&=M^{m+1}(\bm x)-M_s(\bm x)=e_{d}e_{c}^{m-1} \alpha_{M,d} \exp (i \bm{\theta} \cdot \bm{x}).
\end{aligned}
    \right.
    \end{aligned}
\end{equation}
Clearly, if $\bar{\bm{u}}=\bm{u}^{m+1/2}$, then after $m$ steps of evolution, the error will be reduced by $e_c^{m-1}e_g$ times. If, instead, $\bar{\bm{u}}$ taken as the algebraic average over the previous $m-1$ CIS steps, the error will be reduced by $e_c^{m-1}e_d$ times. 

To calculate $e_d$ in a single GSIS step inside the DIG loop, we find that \eqref{GSIS_matrix} remain unchanged, but the source terms are amplified by $A$ times, that is, 
\begin{equation}\label{source_term_DIG}
    \begin{aligned}
        & S_{u,j}={A}\iiint\left[\frac{v_j}{\delta_{rp}}+\frac{\theta_j}{3 \delta_{rp}} \theta_l v_l-2 i \theta_l v_{\langle j} v_{l\rangle}\right] \bar{Y}(e_d,\bm v) d \bm v, \\
        & S_{\tau}={A}\iiint\left[\frac{5}{4 \delta_{rp} } \left(\frac{2}{3} v^2-1\right)-i\theta_lv_l\left(v^2-\frac{5}{2}\right)\right]\bar{Y}(e_d,\bm v) d \bm v,
    \end{aligned}
\end{equation}
where 
\begin{equation}
    A=\frac{1}{m}
    \left[1+\frac{1}{e_c}+\frac{1}{e_c^2}+\cdots+\frac{1}{e_c^{m-1}}\right]=\frac{1-e_c^{-m}}{m(1-e_c^{-1})}
\end{equation}
is a factor characterizing how much larger the average values are compared to those from the last CIS step in a single DIG loop.

\begin{figure}
    \centering
    \includegraphics[width=0.7\linewidth]{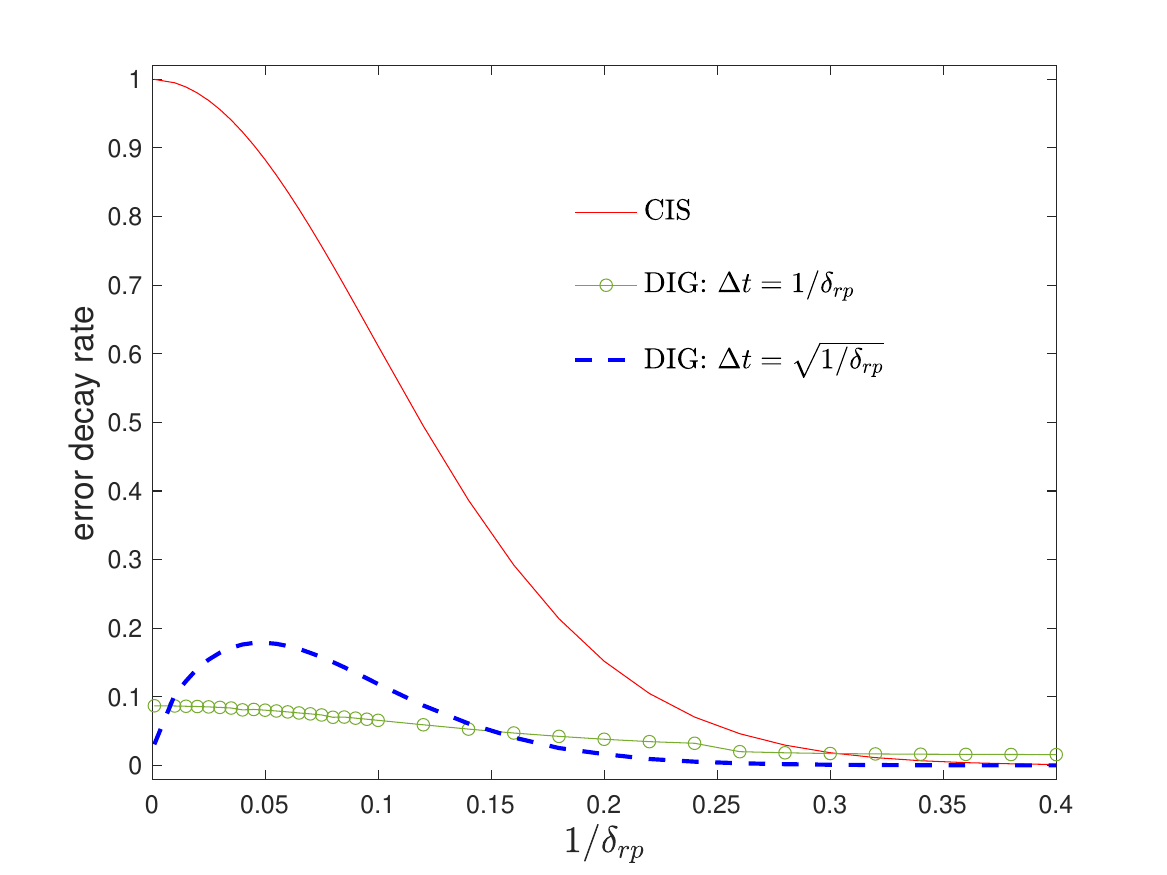}
    \caption{Comparison of error decay rates between the CIS and DIG, when $m=100$. In CIS, the time step is $\delta^{-1}_{rp}$, mimicking that in DSMC. In DIG, the time step are $\delta^{-1}_{rp}$ and $\sqrt{\delta^{-1}_{rp}}$, respectively. Note that the error decay rates in CIS and DIG are calculated i a unit DIG cycle, i.e., $|e_c(\Delta t)|^{100}$ and  $|e_c(\Delta t)|^{99}\times{}|e_d(\Delta t)|$, respectively. }
    \label{fig:DIG}
\end{figure}

Clearly, if $e_c\rightarrow1$, we have $A\approx1$. This means that $e_d\approx{e_g}$.  Thus, the DIG is much faster than the CIS in the continuum limit. For other values of the Knudsen number, Fig.~\ref{fig:DIG} presents a comparison of the convergence behavior between the CIS and DIG when $m = 100$. It should be noted that, in the CIS, the error decay rate is evaluated with $\Delta t = \delta^{-1}_{rp}$. This choice reflects the DSMC convention in which the spatial cell size is on the order of the mean free path, and the time step is on the order of the mean collision time. In contrast, the cell size in DIG can be much larger, and the time step is chosen as $\sqrt{\text{Kn}} \sim \sqrt{\delta^{-1}_{rp}}$, in addition to the standard choice $\Delta t \sim \delta^{-1}_{rp}$. 
Under these circumstances, the DIG remains efficient. Specifically, the error decay rate is below 0.2, implying that after 10 iterations of the DIG, the error can be reduced by approximately seven orders of magnitude, implying that the convergence can be achieved in DSMC after about $1000$ time steps.
When the Knudsen number exceeds 0.4, the CIS  already achieves sufficient efficiency, making the use of DIG unnecessary.

The underlying mechanism for the fast convergence is that the GSIS transforms the hyperbolic kinetic equation into a diffusion-type equation. While the former allows only a finite speed of flow information propagation, the latter exhibits an effectively infinite propagation speed. When solving to steady state, this property greatly accelerates the transmission of flow information across the entire computational domain and significantly reduces unnecessary intermediate iteration steps.


It is interesting to observe from Fig.~\ref{fig:DIG} that the maximum error of DIG obtained with the larger time step is greater than that obtained with the smaller time step. For example, when $\delta^{-1}_{rp}=0.05$ and $\Delta t=\sqrt{\delta^{-1}_{rp}}$, the values of $|e_c|$ and $|e_d|$ $|A|$ are 0.9940 and 0.3238, respectively, whereas for $\Delta t={\delta^{-1}_{rp}}$ are 0.9988, and 0.0911. Consequently, the overall error decay rates in a unit DIG cycle are $|e_c^{99}e_d|=0.1787$ and 0.0806 for $\Delta t=\sqrt{\delta^{-1}_{rp}}$ and $\Delta t={\delta^{-1}_{rp}}$, respectively.
This agrees with the results in Fig.~\ref{fig:CIS_GSIS} that the error decay rate of GSIS, when Kn is small, first increases then decreases when the time step increases.


\section{Analysis of AP property}\label{sec:ap}

We are concerned with whether the NS equation can be derived from \eqref{RTE_steady_d}, or equivalently, whether HoTs in \eqref{GSIS_II_HoTs} tend to \(\mathcal{O}(\text{Kn}^2)\) as \(\text{Kn} \to 0\). If the NS equation can be derived from \eqref{RTE_steady_d}, we say it is an AP-NS scheme.

When the steady state is reached, we have $h^{k+1}=h^k$, then the AP property of CIS and GSIS  in the continuum limit can be analyzed in terms of the following spatial discretization: 
\begin{equation}
  \label{RTE_steady_d}
  \bm{v} \cdot \nabla_{\bm{\delta}} {h} = L_s({h}), \quad\text{with~~}  \nabla_{\bm{\delta}}\left ( \bm{x}, \bm{v}\right ) = \nabla + \Delta x^{n} \bm{\delta} \left ( \bm{x}\right ), 
\end{equation}
where $n$ is the order of spatial accuracy and $\bm{\delta}$ is the spatial derivative.

According to the Chapman-Enskog expansion \cite{CE1970}, the velocity distribution function is expanded as ${h}={h}_0+{h}_1+{h}_2+\cdots$. 
Correspondingly, the stress and heat flux are expressed as ${\sigma}_{l j}=2 \iiint v_{\langle l} v_{j \rangle} ({h}_0+{h}_1+\cdots) d \bm v$ and ${\bm{q}}=\iiint\left(v^2-\frac{5}{2}\right) \bm{v} ({h}_0+{h}_1+\cdots) d \bm v$. However, the conservative variables are only determined by the zeroth-order expansion ${h}_0$ as $[{\rho}, \bm{{u}}, {\tau}] =\iiint \phi(v) {h}_0 d \bm v$, with the compatibility conditions $[0, \bm{0}, 0] =\iiint \phi(v) {h}_n d \bm v, \quad  n=1,2,3,\cdots$. On substituting this expansion into \eqref{RTE_steady_d} and collecting terms of orders $\text{Kn}^{-1}$ and $\text{Kn}^{0}$ gives:
\begin{equation}\label{RTE_steady_d_om1}
\begin{aligned}
 & \text{Kn}^{-1}: \quad 
    {h}_0=\left[
         {\rho}+2 \bm {{u}} \cdot \bm v+{\tau}\left(v^2-\frac{3}{2}\right)
        \right] f_{eq},\\
 & \text{Kn}^{0}: \quad  {h}_1=-{\delta^{-1}_{rp}} 
 \bm{v} \cdot \nabla_{\bm{\delta}}  {h}_0.
\end{aligned}
\end{equation}
Hence the heat flux is approximated as (that for stress can be obtained similarly)
\begin{equation}\label{RTE_steady_d_o0_m1}
  \begin{aligned}
    {\bm{q}} = \int \left(v^2-\frac{5}{2}\right)\bm{v} \left ({h}_0 + {h}_1 \right ) d \bm{v} + \mathcal{O}(\text{Kn}^2)
    &= - {\delta^{-1}_{rp}}{\nabla}_{\bm{\delta}} {\tau} + \mathcal{O}(\text{Kn}^2)\\
    &= - {\delta^{-1}_{rp}}\left ( \nabla + \Delta x^{n}{\bm{\delta}} \right ) {\tau} + \mathcal{O}(\text{Kn}^2).
  \end{aligned}
\end{equation} 
To recover the NS equation, the error term $\Delta x^{n}$ implies that a restriction on grid size $\mathcal{O}(\Delta x^{n}) \sim \mathcal{O}(\text{Kn})$ is imposed for conventional numerical schemes, that is, 
$\Delta x \sim \mathcal{O}(\text{Kn}^{{1}/{n}})$. 
Thus, the conventional transport discretizations have their asymptotic behavior affected by errors at the continuum limit, depending on the precision of the scheme. 

In GSIS, the higher order part of heat flux is extracted as follows
\begin{equation}
  \label{RTE_steady_d_HoT}
  \begin{aligned}[b]
    \text{HoT}_{\bm{q}} = {\bm{q}} + {\delta^{-1}_{rp}} {\nabla}_{\bm{\delta}}{\tau} =  \mathcal{O}(\text{Kn}^2),
  \end{aligned}
\end{equation}
This demonstrates that in GSIS, even when $\Delta{x}\sim \mathcal{O}(1)$, the NS equations are still correctly recovered. That is, although the fluxes (e.g., $\bm{{q}}$) obtained from the BGK equation contain discretization errors of order Kn, these are offset by a compensating negative discretization error in the temperature gradient. Consequently, in the near-continuum limit, solving the BGK equation with GSIS achieves accuracy comparable to directly solving the NS equations.

In the DIG method, after performing $m-1$ iterations of the CIS, the macroscopic synthetic equation \eqref{GSIS_II}, together with the constitutive relation \eqref{GSIS_II_HoTs}, is used to steer the system toward the steady state. However, the above AP analysis based on the steady-state kinetic equation  \eqref{RTE_steady_d} cannot be applied directly. This is because, even if the final GSIS step has already obtained the correct steady-state solution, the subsequent $m-1$ CIS steps would gradually introduce numerical dissipation.
Only if the macroscopic quantities change only slightly during these $m-1$ CIS iterations will the solution produced by DIG be very close to the true steady-state solution. As $\text{Kn}\to0$, the convergence rate of CIS approaches $1 - \mathcal{O}(\text{Kn}^2)$, meaning that after each CIS iteration, the variation in macroscopic quantities is proportional to $\text{Kn}^2$. Therefore, provided that $(m-1)\text{Kn}^2 \ll 1$, the DIG solution will retain the AP-NS property.

It should be emphasized that the above AP analysis, which shows that the NS equation can be reproduced in the near-continuum regime when $\Delta{x}\sim \mathcal{O}(1)$, applies to the bulk region where the variation of physical quantities is small. In regions with steep variations, small cell size may be needed to capture the flow physics. Also, it should be noted that although we say the NS equations can be recovered when $\Delta{x}\sim \mathcal{O}(1)$, this does not imply that the NS equations can be accurately solved at this resolution. Depending on the specific flow problem, a finer spatial resolution may still be required. For example, in the following simulation of Poiseuille flow at Kn=0.01, using only 10 non-uniform cells is insufficient for the Navier–Stokes equations to reproduce the correct velocity profile.

\section{Numerical tests}\label{sec:num}

The DIG algorithm is designed to enhance the efficiency and accuracy of DSMC simulations. In each DIG loop, the standard DSMC method is first executed for $m-1$ steps to obtain averaged macroscopic quantities $\bar{M}$ and the higher-order constitutive relations \eqref{DIG_HoTs}, thereby reducing inherent thermal fluctuations. These are then supplied to the synthetic macroscopic equations \eqref{GSIS_II}, which are solved by the finite-volume method to the steady-state or 500 inner iterations~\cite{DIG}, producing updated macroscopic solutions $M^{m+1}$. The particle distributions in DSMC are subsequently modified to ensure consistency with the $M^{m+1}$ \cite{DIG}; note that for general flows the corresponding nonlinear equations for mass, momentum and energy conservations should be used instead. This procedure is repeated until the overall solution converges to a steady state.
In this section, we choose $m=100$ in a unit DIG cycle and examine how thermal fluctuations affect the convergence.  To this end, the planar Poiseuille flow and the hypersonic flow past a cylinder are considered.

Note that the Boltzmann equation accounts only for gas–gas interactions. To fully describe wall-confined rarefied gas flows, the gas–surface boundary condition must also be specified. The Maxwell diffuse boundary condition is employed here. Upon colliding with the wall, a gas molecule is assumed to fully thermalize with the surface: 
 \begin{equation}
f|_{\bm{v}\cdot{\bm{n}_w}>0} =-\frac{\iiint_{\bm{v}\cdot{\bm{n}_w}<0} (\bm{v}\cdot{\bm{n}_w}) f_{in}d\bm{v}}{\iiint_{\bm{v}\cdot{\bm{n}_w}>0} (\bm{v}\cdot{\bm{n}_w}) \exp\left(-{v^2}/{T_w}\right)d\bm{v}}\exp\left(-\frac{v^2}{T_w}\right),
 \end{equation}
where $\bm{n}_w$ is the wall normal vector pointing to the fluid, $T_w$ is the wall temperature normalized by $T_0$, and $f_{in}$ is the incident distribution function.



In the following DSMC simulations, the variable hard-sphere model is used for argon gas, where the gas viscosity is proportional to the temperature raised to the power $0.81$.

\subsection{Planar Poiseuille flow}

\begin{figure}[t!]  
    \centering
\subfloat[$\text{Kn}=0.01$]{\includegraphics[width=0.45\textwidth,trim=80pt 30pt 120pt 50pt,clip]{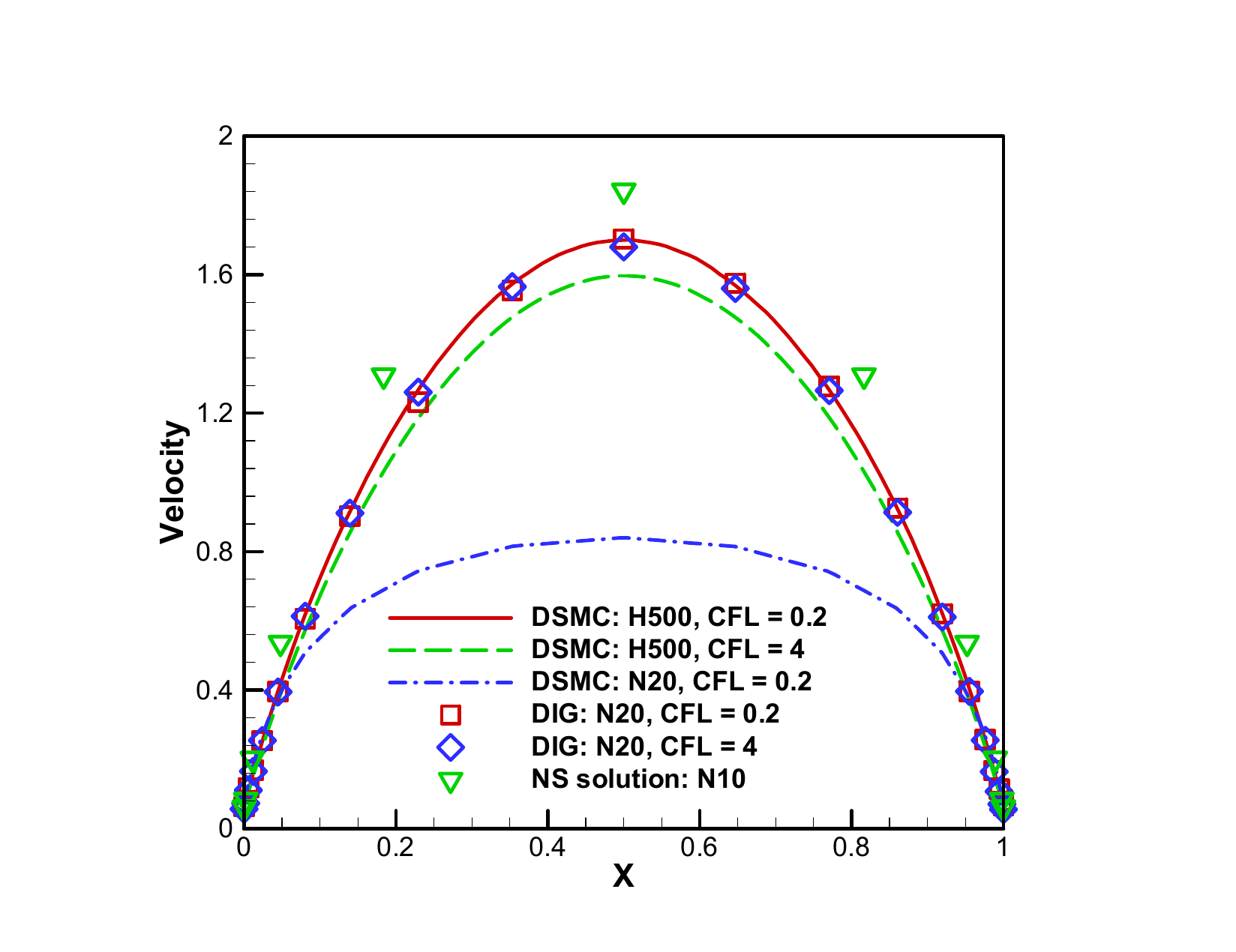}}\\
\subfloat[$\text{Kn}=0.001$]{\includegraphics[width=0.4\textwidth,trim=80pt 30pt 120pt 50pt,clip]{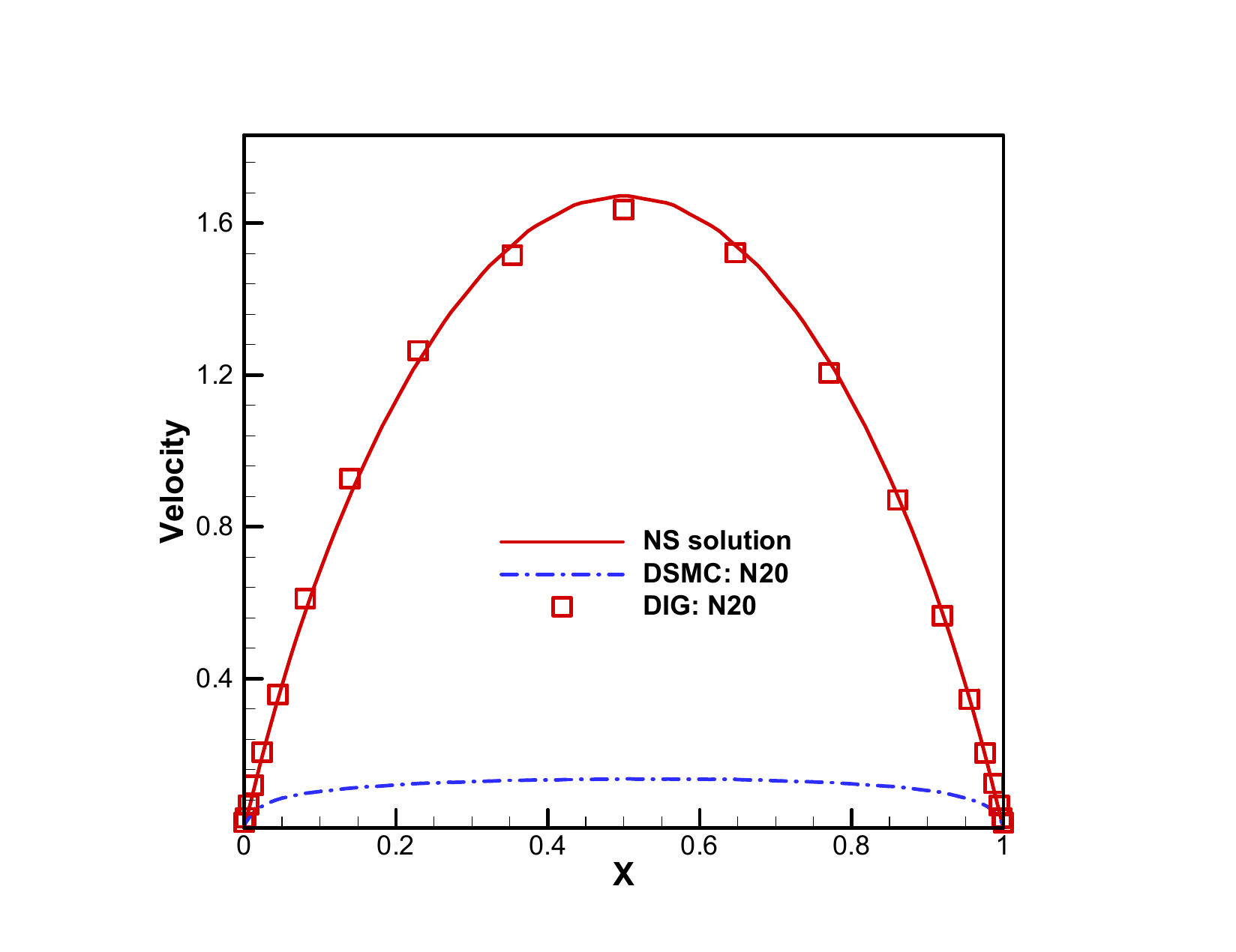}} 
\hspace{0.5cm}
\subfloat[$\text{Kn}=0.1$]{\includegraphics[width=0.4\textwidth,trim=80pt 30pt 120pt 50pt,clip]{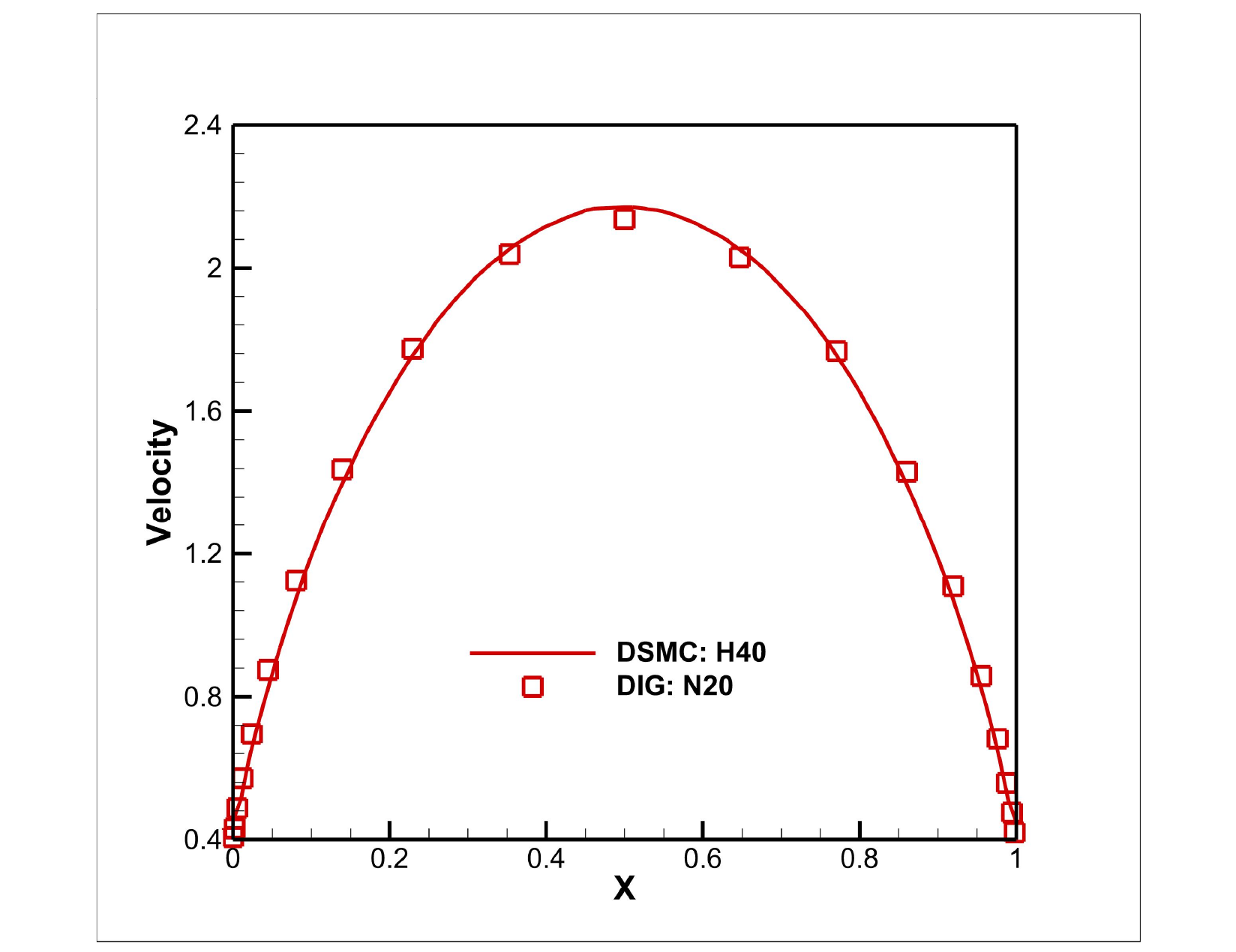}}
    \caption{Velocity profiles in the force-driven Poiseuille flow. H500 and H40 mean that the computational domain is discretized uniformly by 500 and 40 cells, while N20 represents the nonuniform grids generated by \eqref{hyperbolic_tangent_eq}. A CFL number of 0.2 is applied, based on the minimum grid size. }
    \label{Poiseuille_figure}
\end{figure}

We simulate the one-dimensional force-driven Poiseuille flow to assess the accuracy and efficiency of the DIG method, as this flow is highly sensitive to numerical dissipation when a large cell size is employed \cite{Wang2018Comparative}. The flow is confined between two infinite plates maintained at a temperature $T_0$, positioned at $x_1=0$ and $x_1=L$. The gas molecules are subjected to an external force in the $x_2$ direction, where the normalized acceleration is $10\text{Kn}$. Initially, 200 simulated particles per cell are sampled from a Maxwell distribution corresponding to unit temperature and zero velocity. 

To capture the Knudsen layer in the vicinity of solid walls, a non-uniform grid is employed in the $x_1$ direction, using a symmetric hyperbolic–tangent stretching function:
\begin{equation}
    \begin{aligned}
        \frac{x_i}{L}=\frac{1}{2}+\frac{\tanh \left[\theta\left(\frac{2 i}{N-1}-1\right)\right]}{2 \tanh (\theta)}, \quad i=0,1, \ldots, N-1.
    \end{aligned}\label{hyperbolic_tangent_eq}
\end{equation}
When $N=21$ and $\theta=3.01$, the largest and smallest cell sizes—located at the walls and at the center of the domain, respectively—are 0.147 and 0.002.

Figure \ref{Poiseuille_figure}(a) shows the numerical results for different cell sizes when Kn=0.01. First, although in DIG the NS equations can be derived from the kinetic equation with $\Delta x\sim \mathcal{O}(1)$, using only 10 non-uniform cells is not sufficient for accurately solving the NS equations. Instead, about 20 cells are required. 
However, in traditional DSMC, roughly 500 uniform cells are required, and the CFL number must be kept small—for example, around 0.2. If the CFL number is increased to 4 or the cell size is enlarged to five times the mean free path, numerical dissipation \cite{Garcia1998, Garcia2000} emerges, leading to an underestimation of the velocity profile when a uniform spatial discretization is used. In contrast, DIG can still deliver accurate results with only 20 uniform cells (where the largest cell size is about 15 times the mean free path), regardless of whether the CFL number is 0.2 or 4. 

\begin{figure}[t!]
\centering
     \includegraphics[width=0.32\textwidth, trim=80pt 40pt 120pt 50pt,clip]{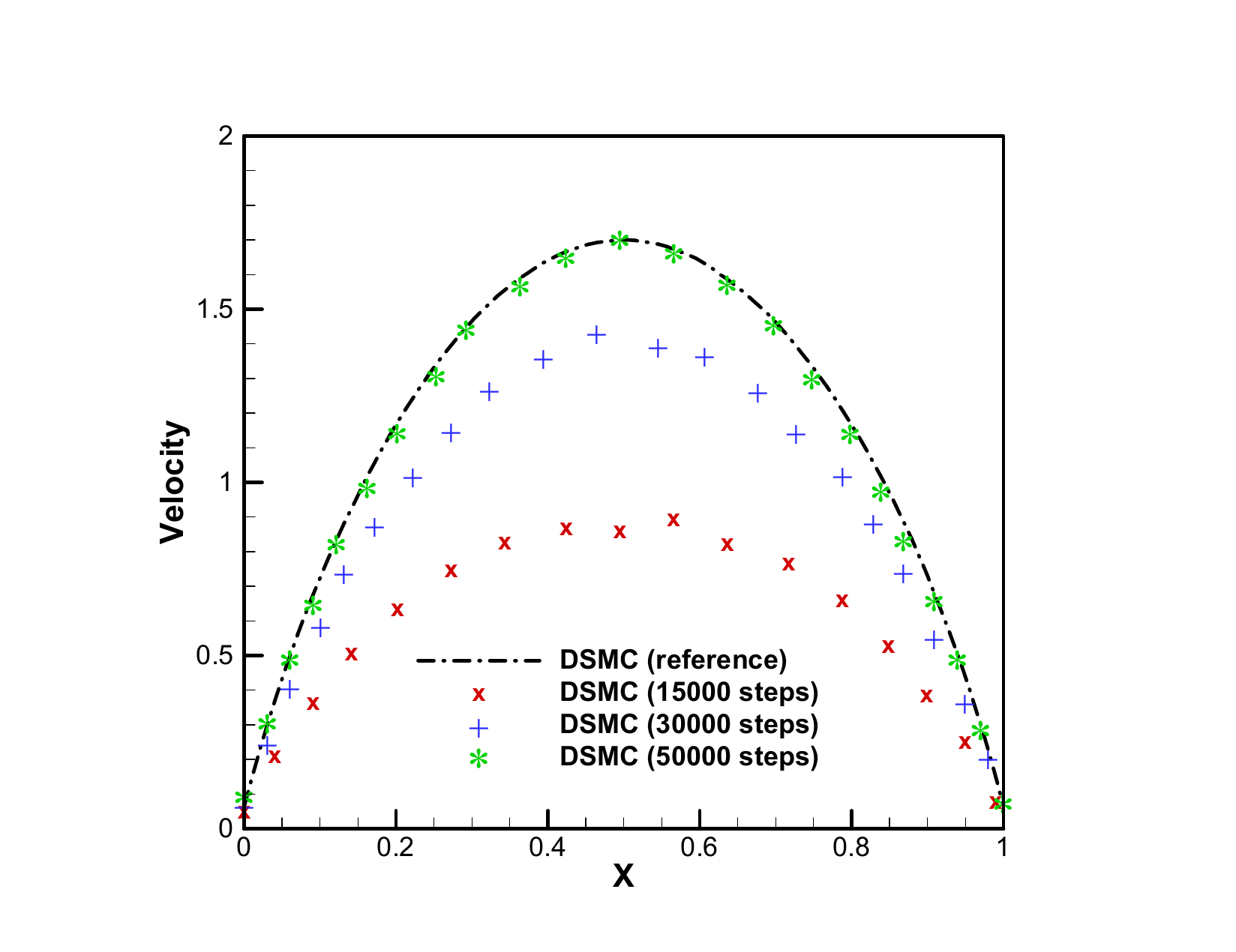}
    \includegraphics[width=0.32\textwidth, trim=80pt 40pt 120pt 50pt,clip]{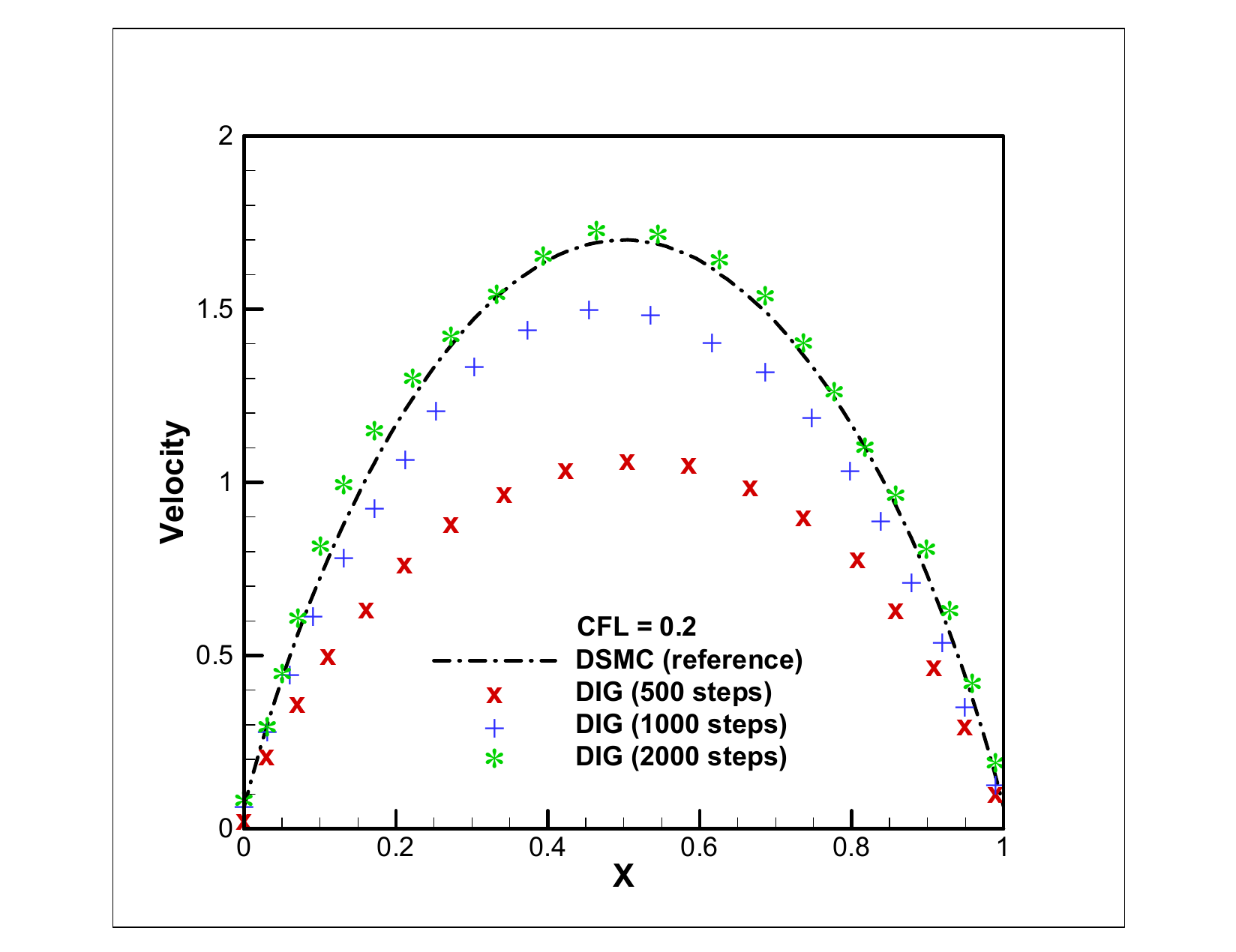}
    \includegraphics[width=0.32\textwidth, trim=80pt 40pt 120pt 50pt,clip]{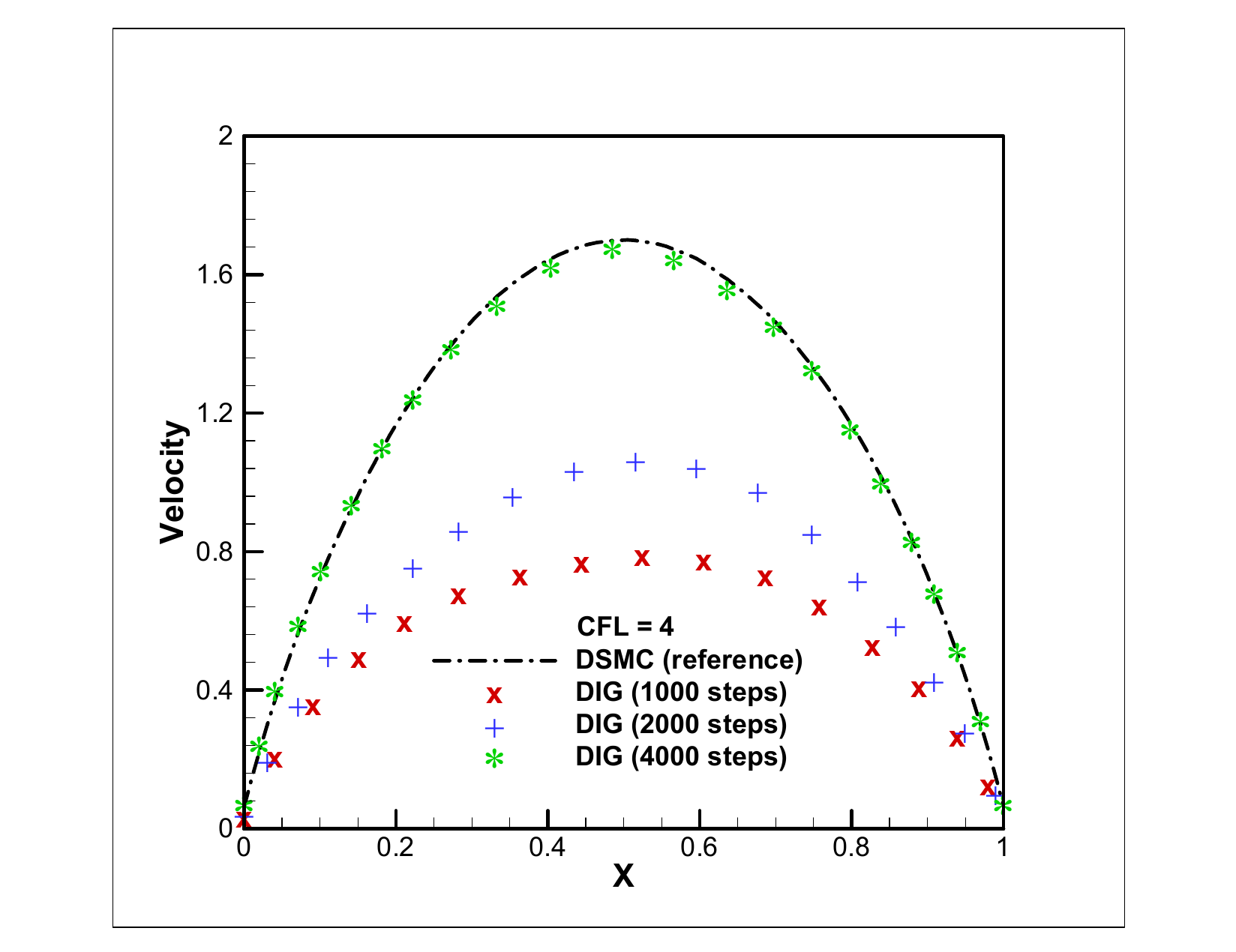}
    \caption{Convergence history of velocity profiles at Kn = 0.01. 500 uniform and 20 non-uniform cells are used in DSMC and DIG, respectively. The CFL number in DSMC is  0.2, while that in DIG is 0.2 and 4. Therefore, the time step in the left two figures is the same. For each line, the result is obtained by averaging over the previous 100 time steps.
    }
    \label{Poiseuille_evolution_velocity}
\end{figure}

Figure \ref{Poiseuille_figure}(b,c) further shows the velocity profiles of Poiseuille flow obtained by the DSMC and DIG methods at $\text{Kn}=0.001$  and 0.1, respectively. At Kn = 0.1, the DIG method produces velocity profiles that agree well with those of DSMC, using the same order of cell numbers. When the Knudsen number decreases to 0.001, DSMC requires some 5000 uniform cells and an extremely large number of iterations (because the error decays at 
$1-\mathcal{O}(\text{Kn}^2)$, roughly $1/\text{Kn}^2=1,000,000$ iterations are needed) to reach a steady state before sampling the profiles for reference data. Therefore, the DSMC results for this case are not presented; instead, the NS solution is provided for comparison. The DIG method, benefiting from its AP–NS properties, accurately captures the flow profiles using only 20 non-uniform cells, where the largest cell size is about 150 times the mean free path. If the traditional DSMC method uses this non-uniform mesh, the resulting velocity profile is roughly an order of magnitude smaller, indicating severe numerical dissipation.

Figure \ref{Poiseuille_evolution_velocity} illustrates the convergence history of velocity at Kn = 0.01. When the same time step is used, DSMC requires approximately 50,000 iterations and 790 seconds to reach steady state, whereas DIG converges in only 2,000 iterations and 2 seconds (all DSMC and DIG results are generated using our in-house unstructured-mesh solver, executed on a single CPU core). This corresponds to roughly 25-fold and 400-fold reductions in iteration count and computational time, respectively. Moreover, the smaller the Knudsen number, the greater the reduction in DIG’s computational time relative to the traditional DSMC. 

A CFL number of 4 is also tested in the DIG method. As shown in Fig.~\ref{Poiseuille_evolution_velocity}, the correct velocity profile can still be recovered, although more DIG loops are required. This is consistent with the Fourier stability analysis in Fig.~\ref{fig:DIG}, which shows that a larger time step yields a larger maximum error decay rate and therefore requires more iteration steps. For this reason, it is recommended that the time step in DIG follow the traditional DSMC choice. This also offers the added benefit of reducing the computational time for particle movement in unstructured cells.

\subsection{Shock wave passing over square cylinder}

\begin{figure}[t!]
\centering
   {\includegraphics[width=0.45\textwidth, trim=10pt 20pt 70pt 110pt,clip]{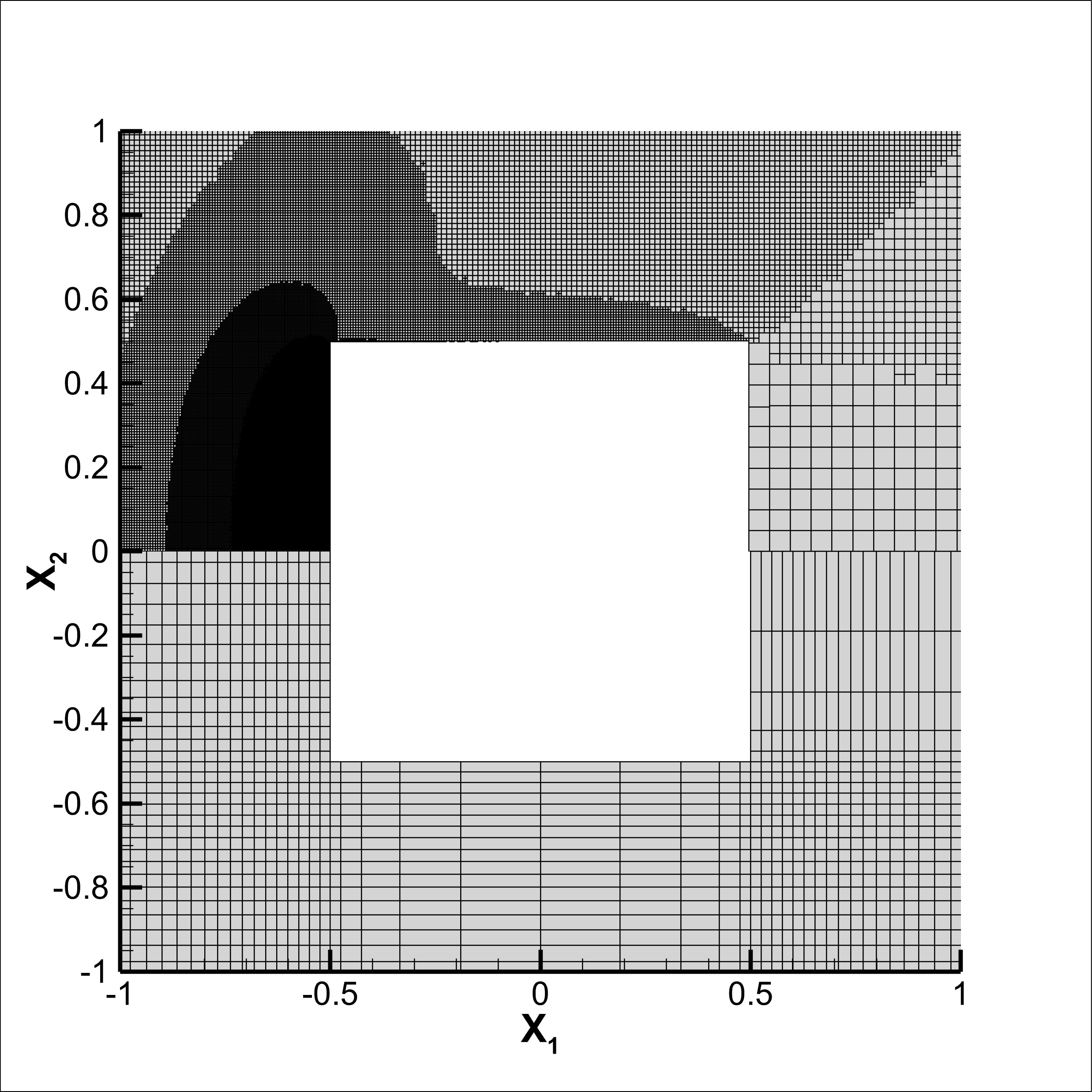}}
    \hspace{0.5cm}
   {\includegraphics[width=0.45\textwidth, trim=10pt 20pt 70pt 110pt,clip]{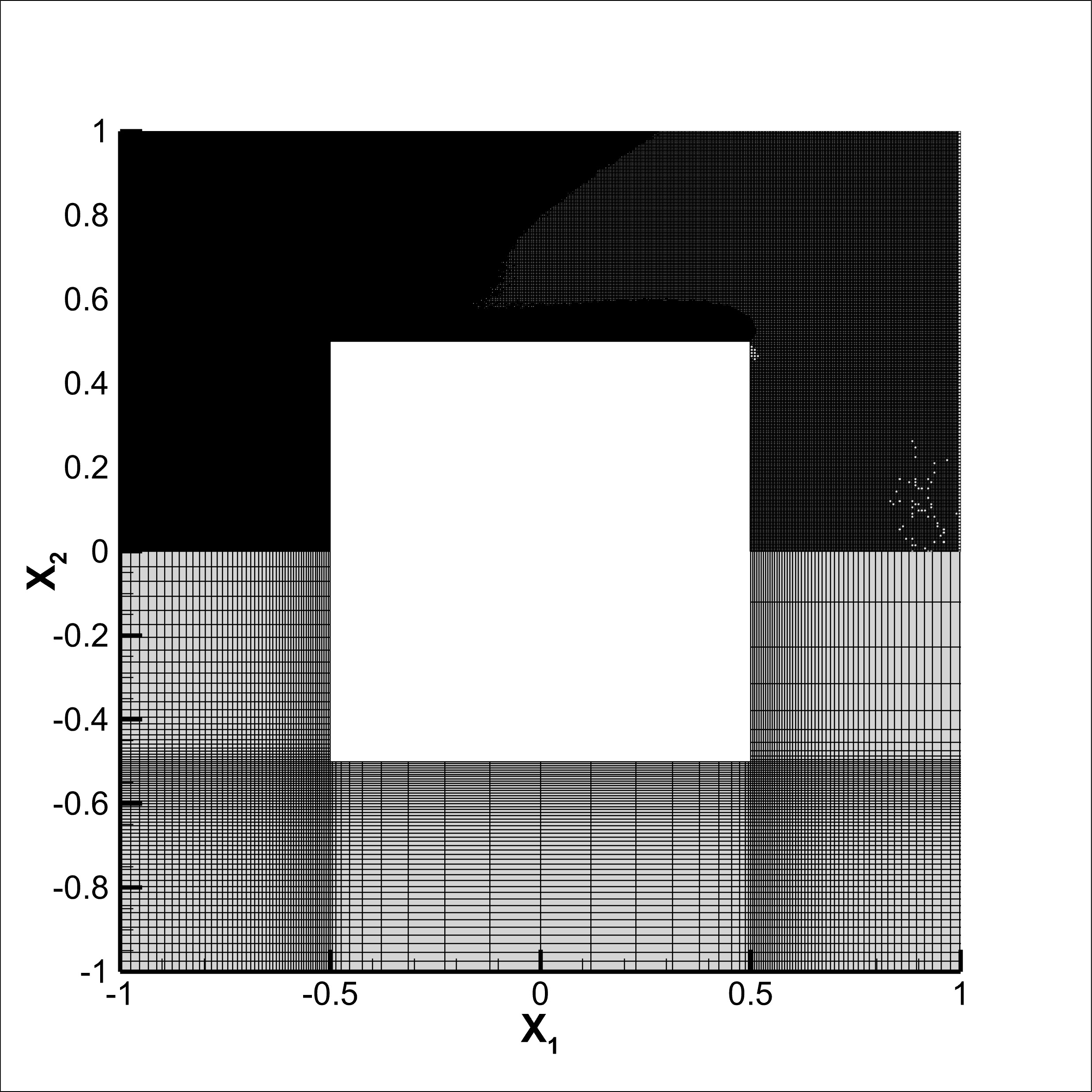}}
    \\
    {\includegraphics[width=0.42\textwidth, trim=10pt 50pt 270pt 40pt,clip]{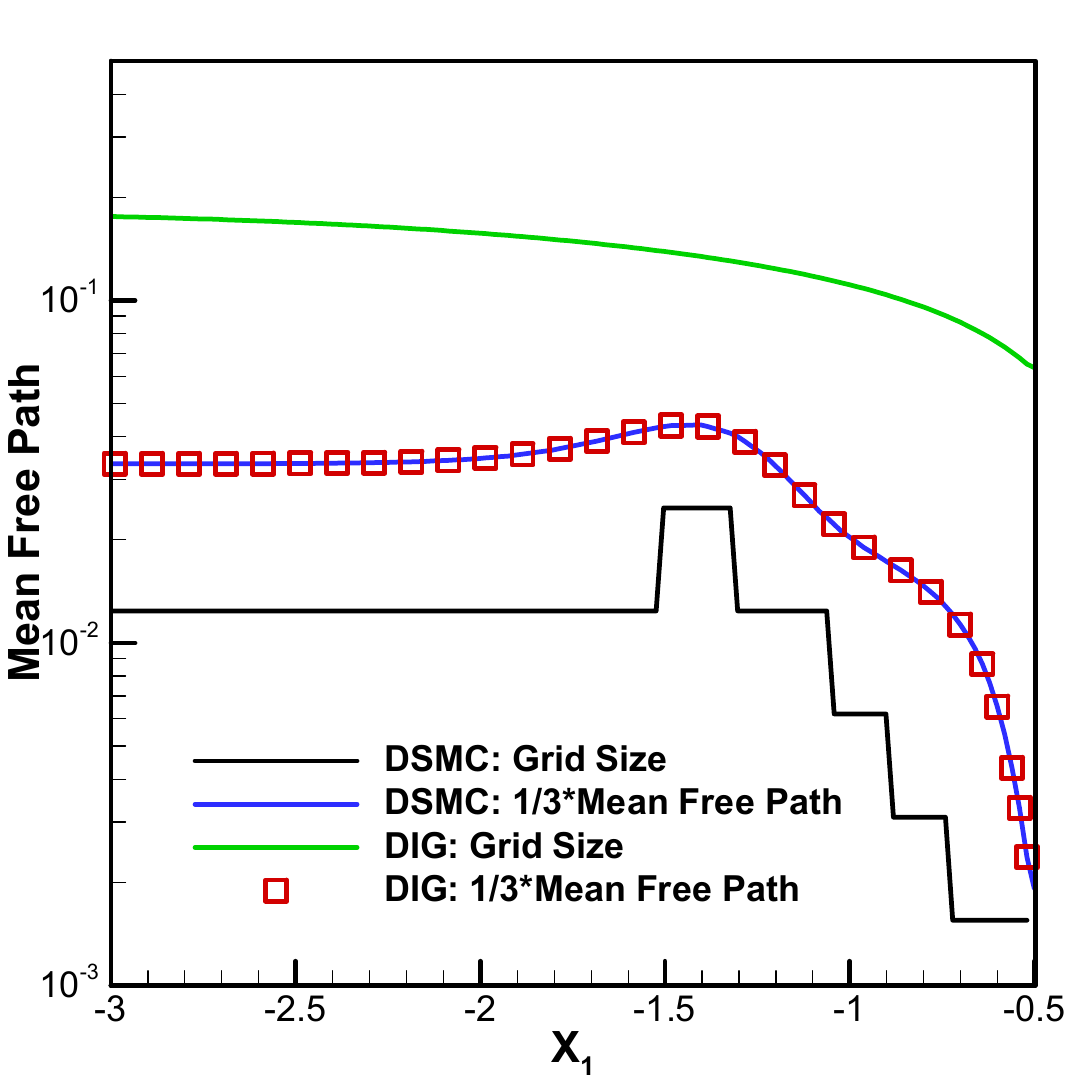}}
    \hspace{0.5cm}
   {\includegraphics[width=0.42\textwidth, trim=10pt 50pt 270pt 40pt,clip]{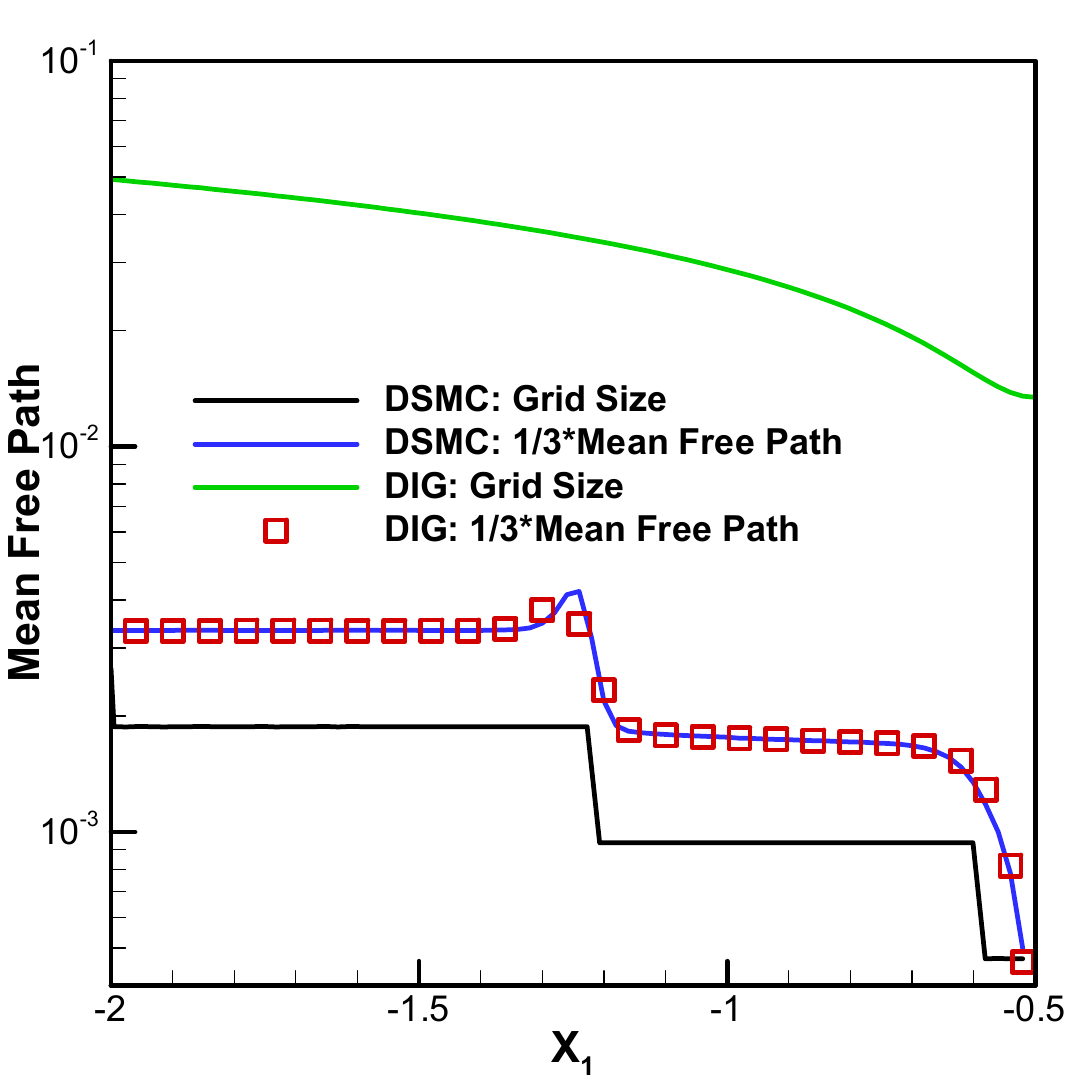}}
    \caption{First row: computational grids in DSMC (upper half) and DIG (lower half) near the square cylinder. In the open-source DSMC code SPARTA, DSMC requires about 838,545 cells at Kn = 0.1 and 5,814,370 cells at Kn = 0.01, with only 12,750 and 51,000 cells for DIG, respectively.
    Second row: comparison of the mean free path and grid size between DIG and DSMC. The mean free path first increases and then decreases when approaching the left side of the cylinder because, due to strong compression, the temperature rises first, but is followed by a significant increase in density. The Knudsen numbers are 0.1 and 0.01 in the left and right columns, respectively.
    }
    \label{square_cylinder_Gridsize}
\end{figure}

In force-driven flow, the mean free path remains constant throughout the computational domain once the Knudsen number is specified. Here, we investigate the shock-wave interaction with a square cylinder, where the local mean free path varies significantly across the computational domain. That is, the gas density becomes highly compressed at the cylinder’s leading edge and greatly reduced at its trailing edge, due to the flow compression and expansion, respectively.

The computational domain consists of an outer square boundary with a side length of $11L$ and a concentric inner square representing a solid square surface with a side length of $L$. The inner square surface is maintained at a normalized wall temperature of $ T_w = 1$. The isothermal walls are modeled with the diffuse reflection boundary condition.  
The freestream number density, temperature, and the side length of the inner cylinder are chosen to be the reference values. 
The left and right far-field boundaries are specified as inlet boundaries maintained in equilibrium, where the normalized density and temperature are 1, and flow velocity is $(\sqrt{6/5}\text{Ma}_{\infty},0,0)$, being the free stream Mach number $\text{Ma}_{\infty}=5$.

Two Knudsen numbers in the far field, Kn = 0.1 and Kn = 0.01, are considered. As shown in Fig.~\ref{square_cylinder_Gridsize}, non-uniform Cartesian grids are used in DIG, according to \eqref{hyperbolic_tangent_eq} in the $x_1$ and $x_2$ directions: when Kn = 0.1, the cylinder surface is discretized using 55 points (25 at the windward side and 10 on others; corresponding to the parameter of $\theta=0.932$ and 1.895, respectively) with a minimum spacing of 0.025. In the direction perpendicular to the square surface,  100 layers of non-uniform outer grids with $\theta=2.264$ are generated, say, over the range $-10.5\le{}x_1\le-0.5$; then only the grids for $x_1\in[-5.5,-0.5]$ are used. When Kn = 0.01, the surface discrete points increase to 110 points (50 at the windward side and 20 others; with $\theta=1.667$ and 2.406, respectively) with a minimum spacing of 0.005. In the direction perpendicular to the square surface, 200 layers of non-uniform outer grids with $\theta=1.654$ are generated over the range $-10.5\le{}x_1\le-0.5$, of which only half are used in the simulation.


\begin{figure}[t]
    \centering
     \includegraphics[width=0.42\textwidth,trim = 10pt 50pt 250pt 50pt,clip]{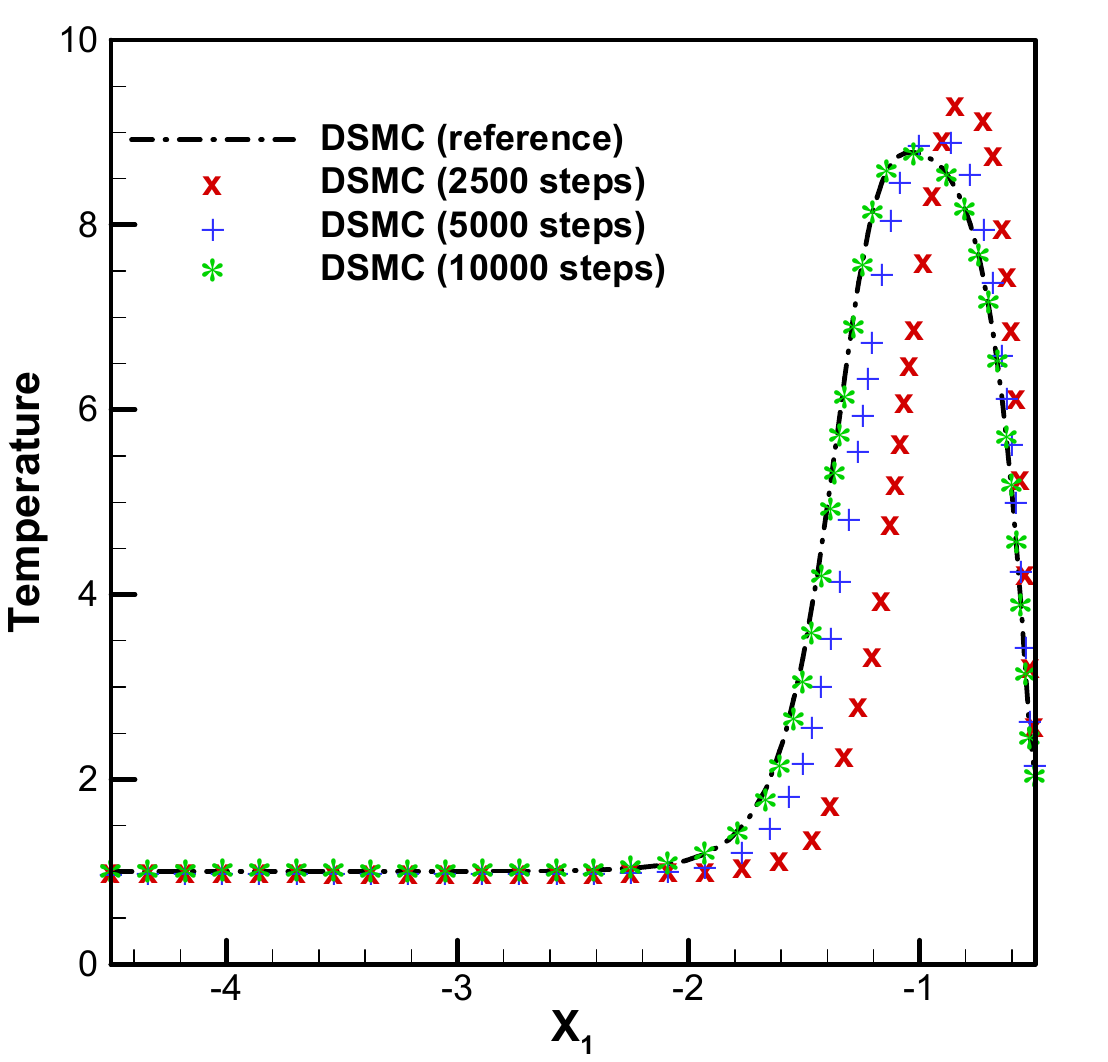}
    \includegraphics[width=0.42\textwidth,trim = 10pt 50pt 250pt 50pt,clip]{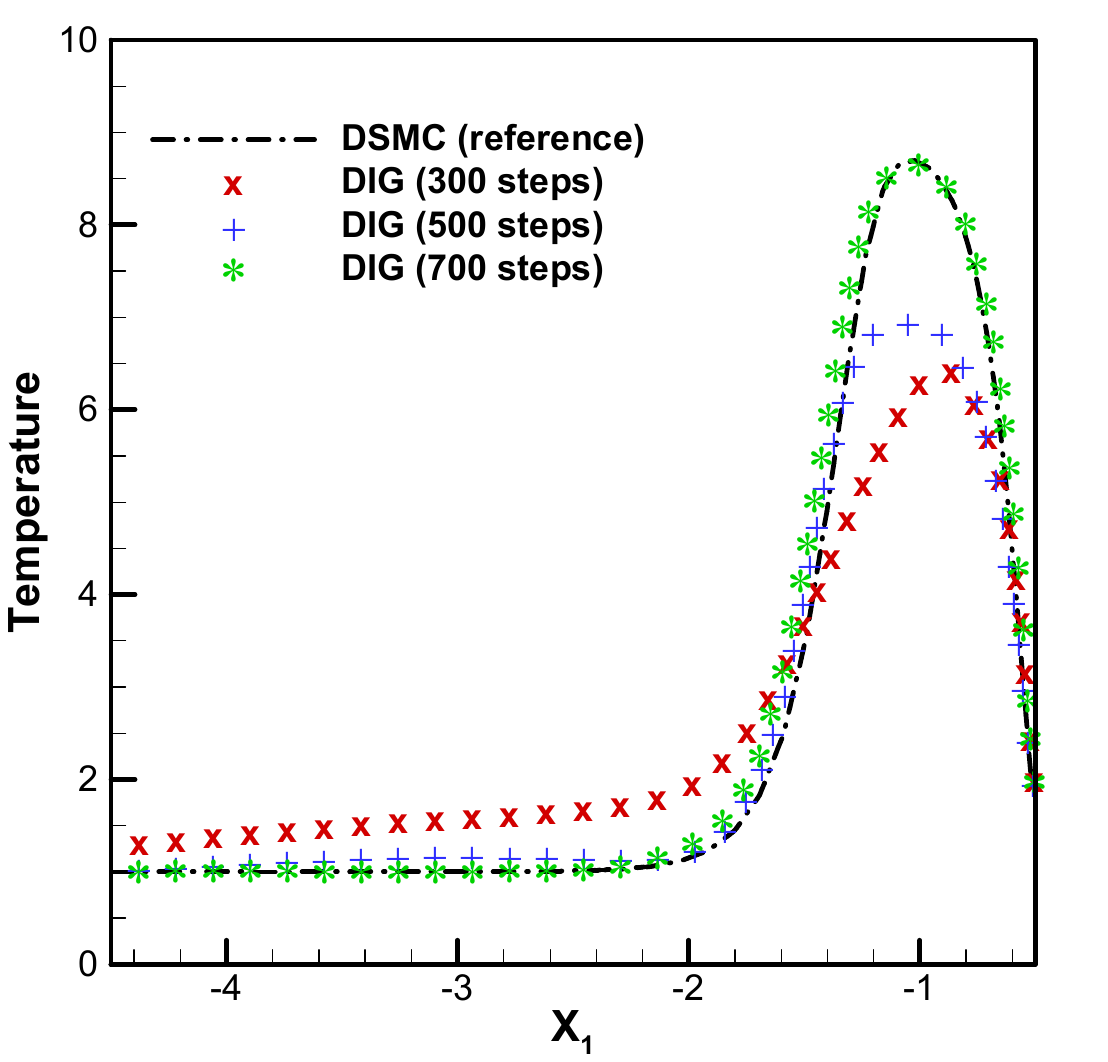}\\
    \includegraphics[width=0.42\textwidth,trim = 10pt 50pt 250pt 50pt,clip]{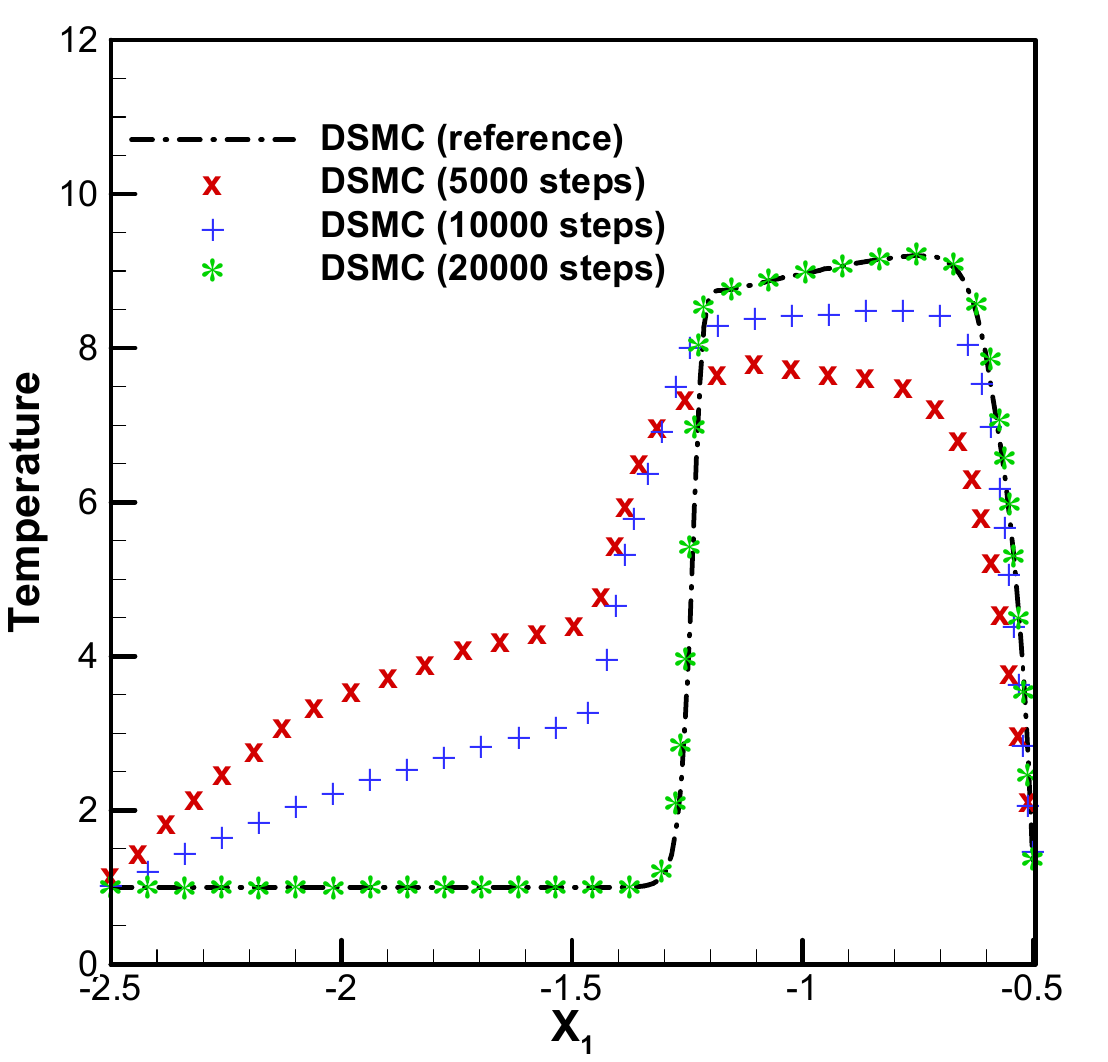}
    \includegraphics[width=0.42\textwidth,trim = 10pt 50pt 250pt 50pt,clip]{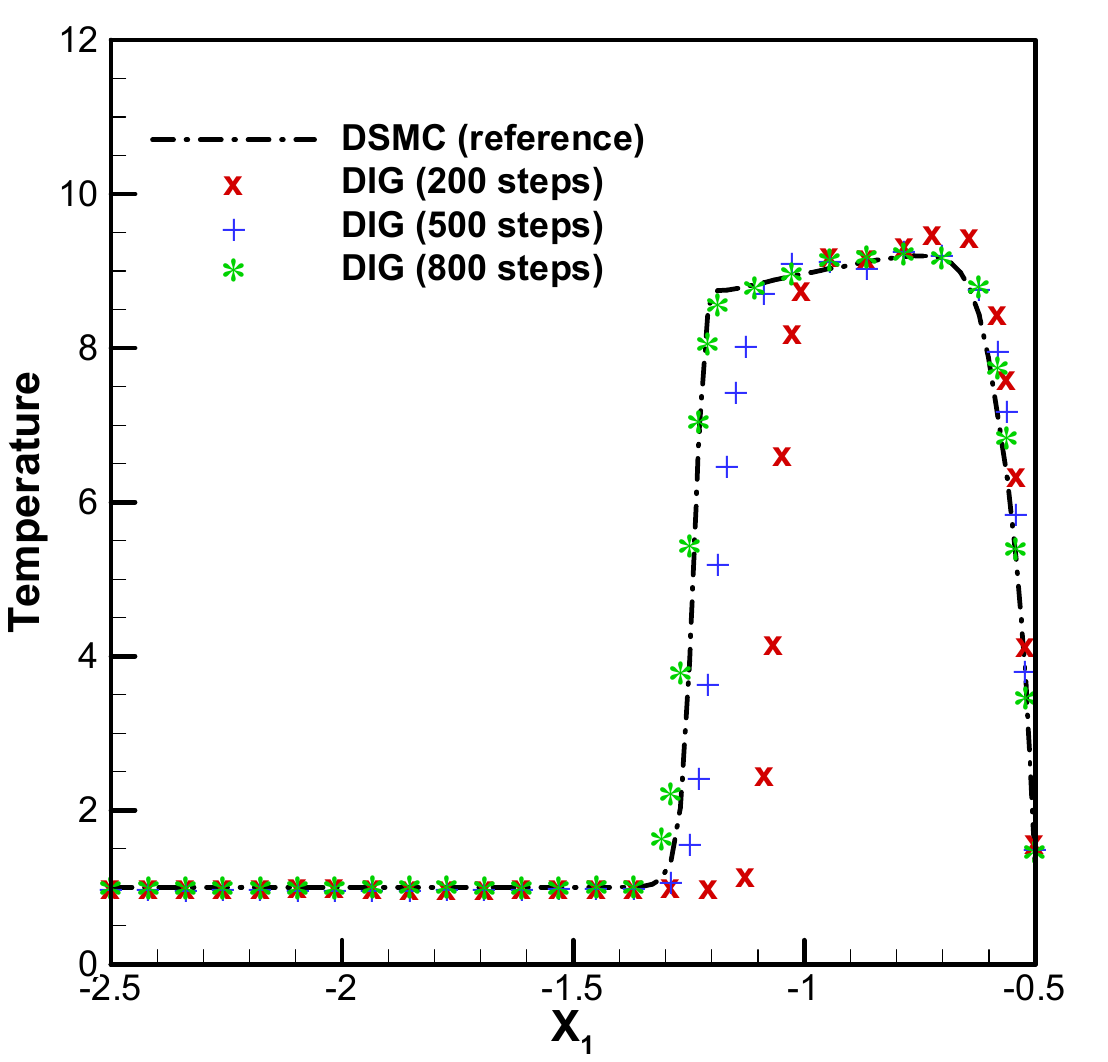}
    \caption{Comparison of the convergence history of the temperature along the stagnation line between the DSMC and DIG, when Kn = 0.1 (first row) and 0.01 (second row). }
    \label{fig: stagnationline_evolution_T}
\end{figure}

Figure \ref{fig: stagnationline_evolution_T} compares the convergence history of the temperature along the stagnation line between the original DSMC and DIG, where the CFL numbers are both 0.2. At Kn = 0.1, DIG reaches a steady state in about 800 iterations—roughly an order of magnitude fewer than required by DSMC. At Kn = 0.01, the computational advantage of DIG becomes even more pronounced, achieving convergence in only 700 iterations and providing an acceleration factor of 25 relative to DSMC. These results clearly demonstrate the fast-converging property of the DIG method.

\begin{figure}[t!]
\centering
\vspace{-1.5mm}
\includegraphics[width=0.45\textwidth,trim=20pt 40pt 250pt 30pt,clip]{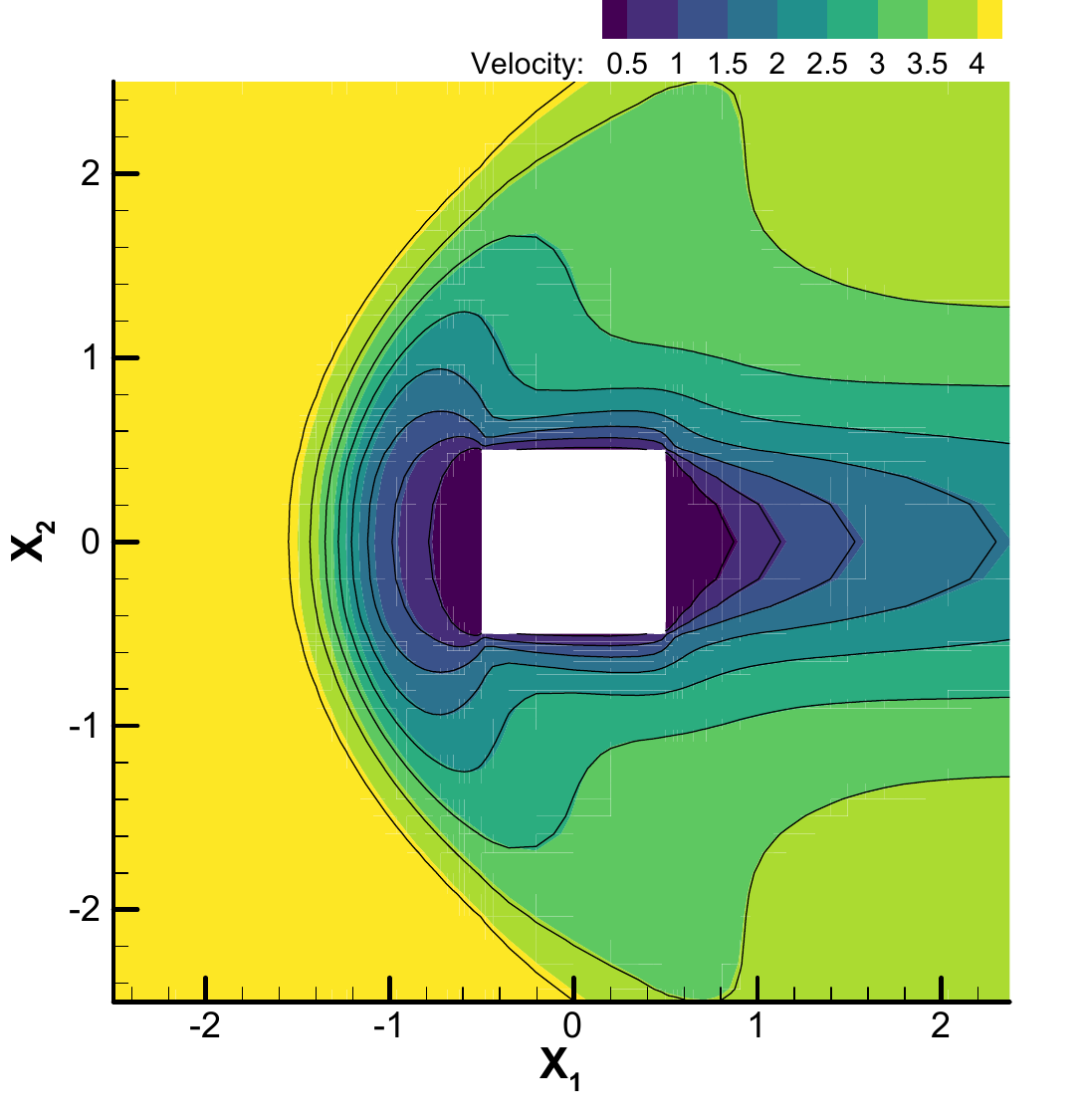}
\includegraphics[width=0.45\textwidth,trim=20pt 40pt 250pt 30pt,clip]{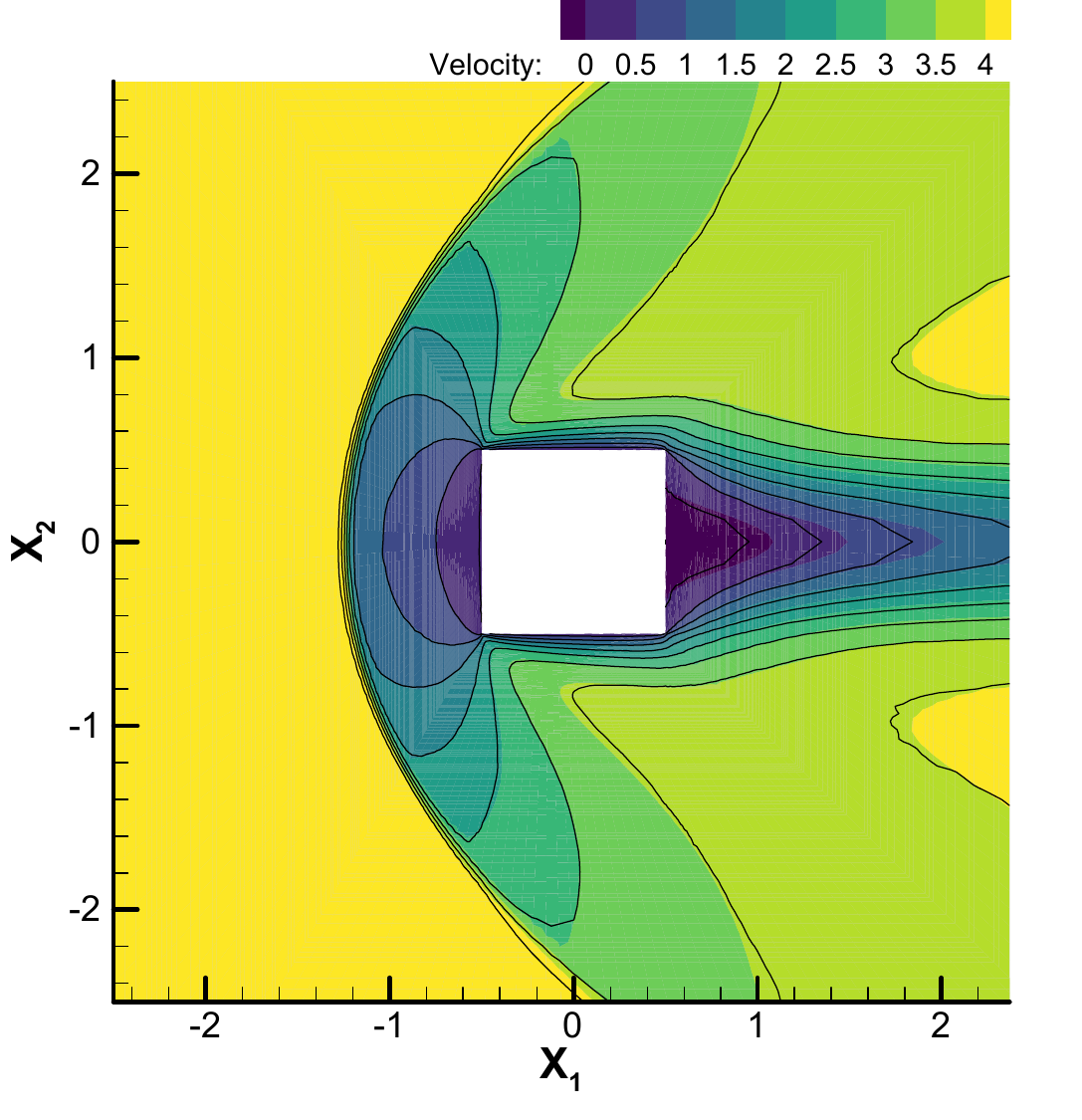}
\\
\includegraphics[width=0.45\textwidth,trim=20pt 40pt 250pt 20pt,clip]{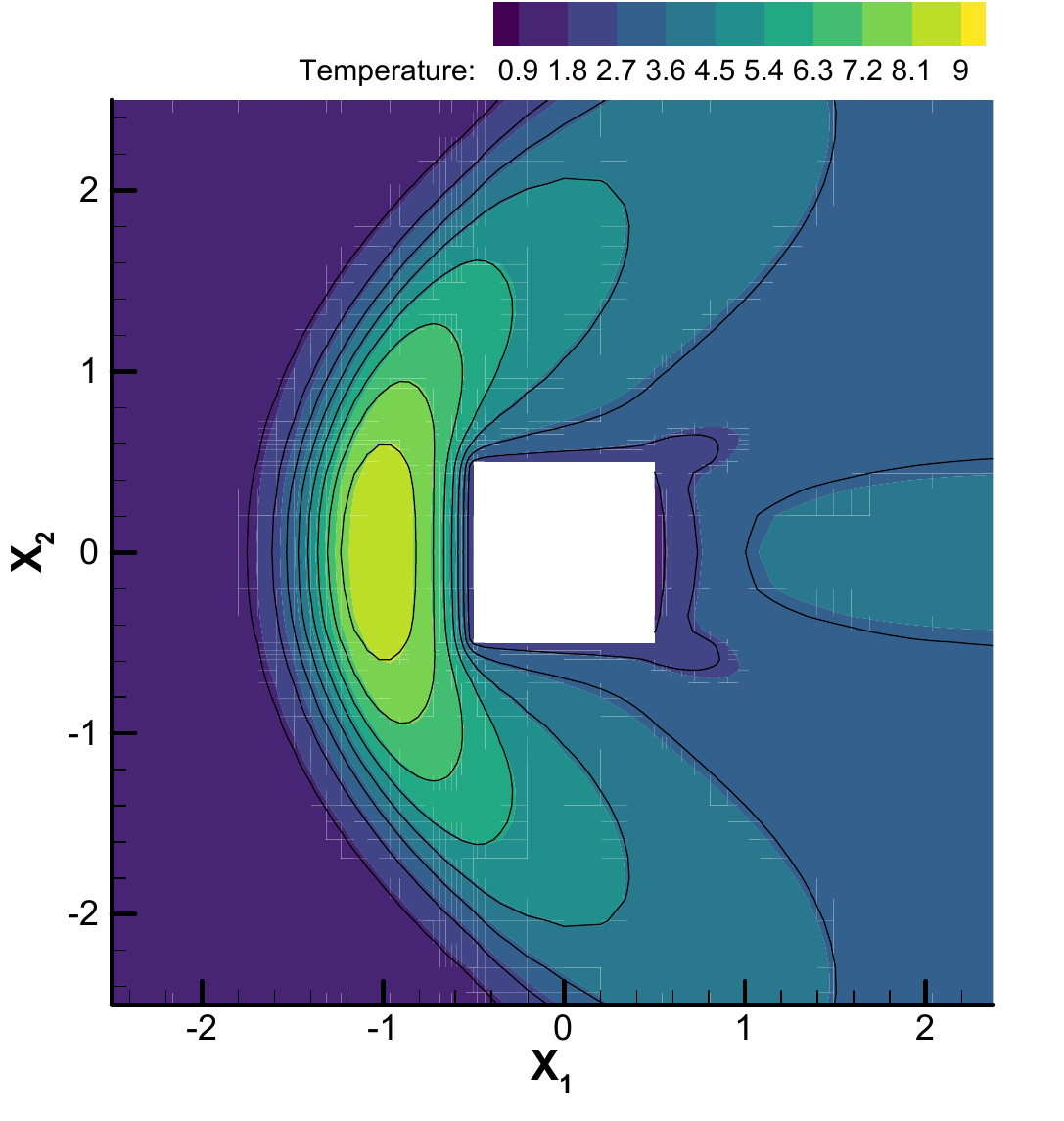}
\includegraphics[width=0.45\textwidth,trim=20pt 40pt 250pt 20pt,clip]{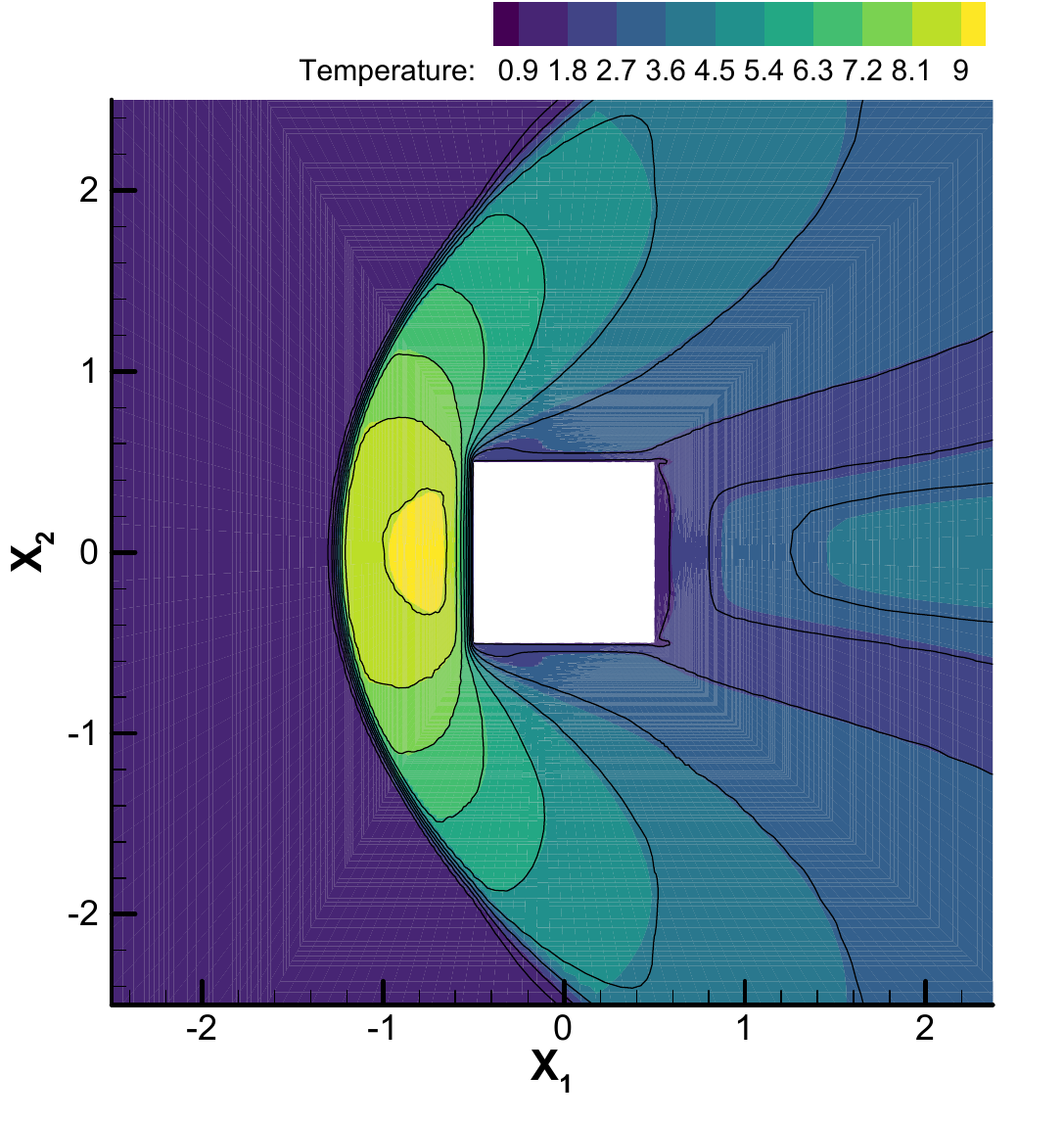}
\\
\includegraphics[width=0.32\textwidth, trim=10pt 50pt 280pt 60pt,clip]{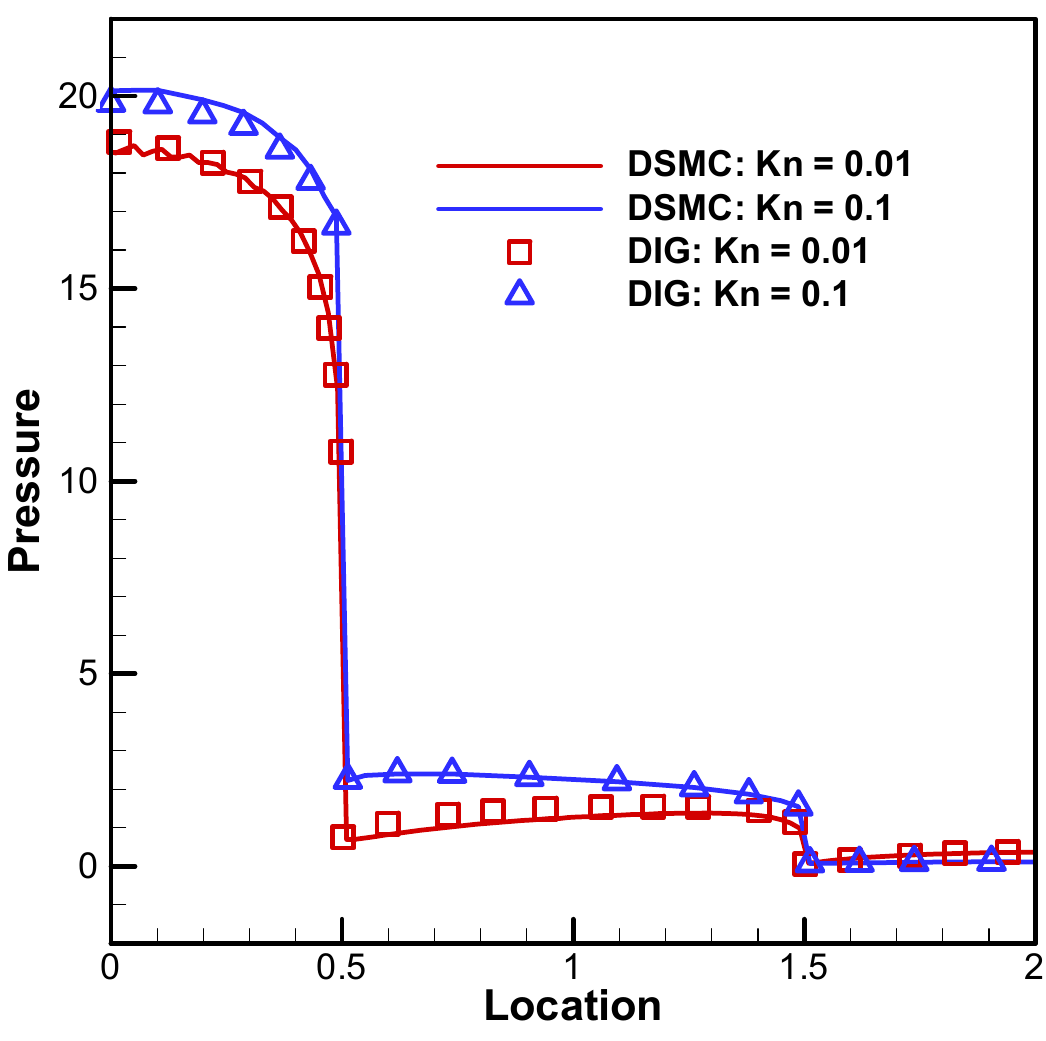}
\includegraphics[width=0.32\textwidth, trim=10pt 50pt 280pt 60pt,clip]{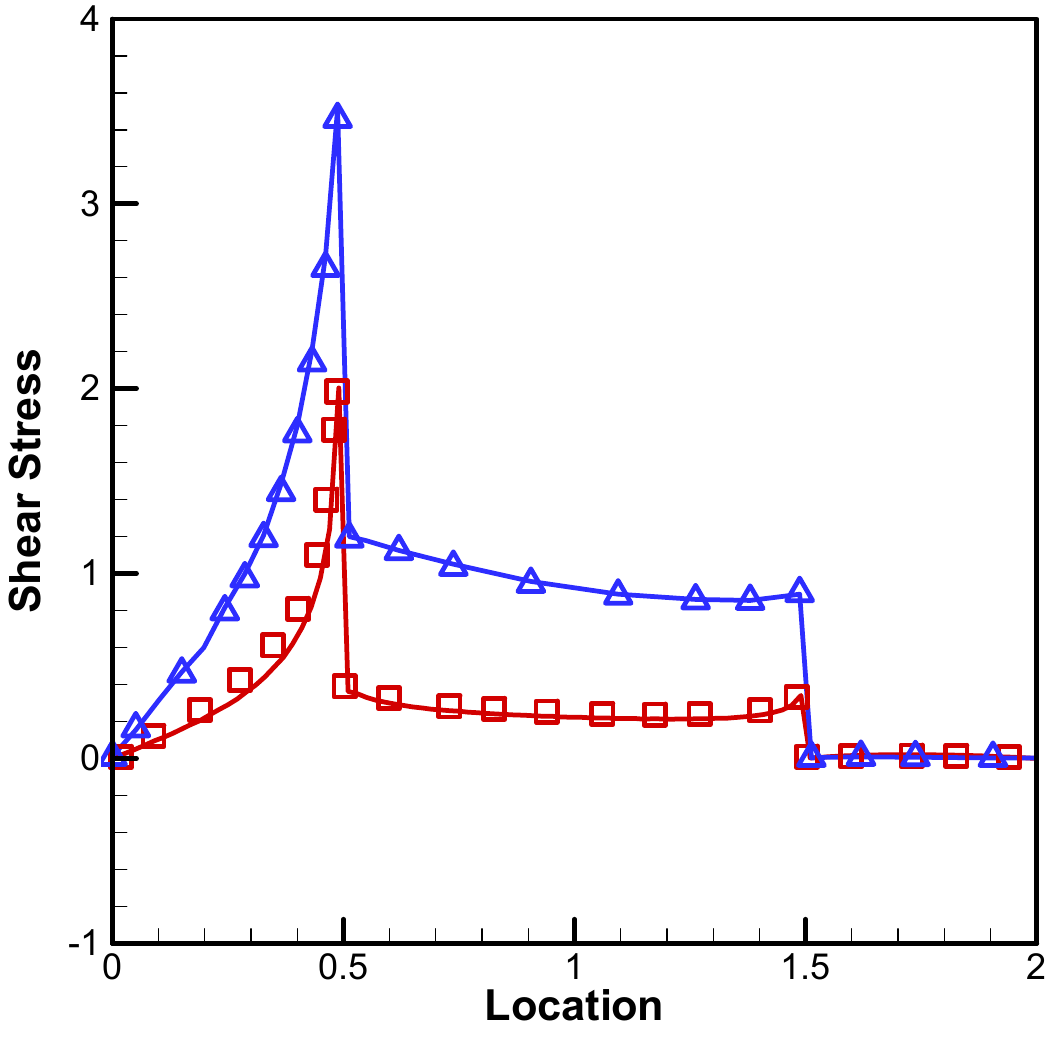}
\includegraphics[width=0.32\textwidth, trim=10pt 50pt 280pt 60pt,clip]{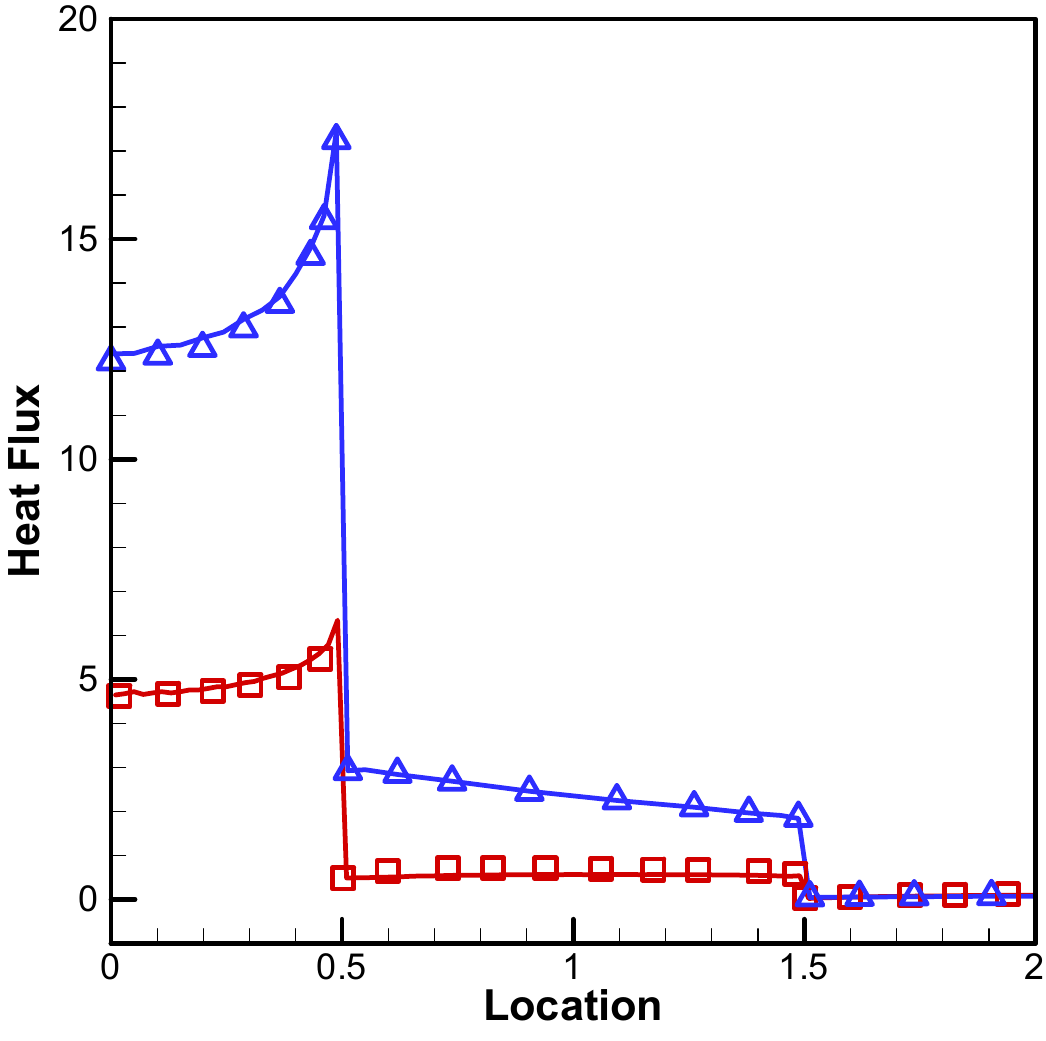}
\caption{First and second rows: contours of velocity and temperatures. 
DIG and DSMC results are shown as solid black lines and the colored background, respectively. The Knudsen numbers in the left and right columns are Kn = 0.1 and 0.01, respectively. Third row: the pressure, shear stress, and heat flux on the surface of the cylinder, where the location is the surface distance measured from the midpoint of the windward of the cylinder.}
\label{fig:Contour_Ma5_square_macro}
\end{figure}

Figure~\ref{fig:Contour_Ma5_square_macro} further compares the steady-state velocity and temperature contours obtained by DSMC and DIG. The DIG results show good agreement with DSMC, even though the number of grid cells is reduced by two orders of magnitude. The pressure, shear stress, and heat flux along the cylinder surface are also compared. Excellent agreement is exhibited in all presented results for both methods. These clearly demonstrates the AP-NS property of the DIG method.

\section{Conclusions}\label{sec:conclusion}

In summary, using a divide-and-conquer strategy, we analyzed the fast-converging and asymptotic-preserving properties of the direct intermittent general synthetic iterative scheme (DIG). First, through Fourier stability analysis, we computed the error decay rates of DIG by accounting for the effects of time discretization while keeping the spatial derivatives intact in the linearized BGK kinetic equation. It is found that while the conventional iteration scheme exhibits an error-decay rate close to 1 in the near-continuum regime, the maximum error-decay rate of DIG is below 0.2, indicating that after one cycle of DIG (e.g., 100 DSMC evolution steps), the error is reduced by approximately a factor of 5. 
Second, using the Chapman–Enskog expansion, we demonstrated that DIG possesses the AP-NS property in the steady-state regime by accounting for spatial discretization errors while neglecting the time derivative in the linearized BGK kinetic equation; specifically, it recovers the NS equations when the cell size satisfies $\Delta x \sim \mathcal{O}(1)$, rather than being constrained by the mean free path. 
Third, two challenging numerical tests of DSMC simulations accelerated by DIG demonstrated that DIG helps bypass unnecessary intermediate evolutions and reduces the number of spatial cells required, thereby reducing computational memory usage and runtime by several orders of magnitude in near-continuum flows.

We note that the moment-guided DSMC method \cite{Degond2011} was proposed to reduce statistical variance by coupling the DSMC algorithm with a set of macroscopic (moment) equations. This approach is conceptually similar to the DIG method in the sense that both kinetic and macroscopic equations are solved, but they differ in the formulation of constitutive relations and in the numerical treatment of the moment equations. To be specific, in the moment-guided DSMC method~\cite{Degond2011}, the stress and heat flux are directly extracted from the DSMC simulation, rather than evaluated through \eqref{RTE_steady_d_HoT} to remove the numerical dissipation due to spatial discretization. Consequently, the method does not possess the AP-NS property, although it helps to reduce the thermal fluctuation. In addition, since the macroscopic equations are solved only over a single time step rather than iterated to a steady state, the scheme lacks the fast-converging property. A similar deficiency exists in the NS–DSMC coupling method~\cite{schwartzentruber-2006, schwartzentruber-2007}.

In addition to its fast-converging and AP properties, DIG has the potential to be directly applied to flow simulations involving complex physical and chemical processes, as it retains the inherent flexibility of DSMC and benefits from sophisticated numerical methods available for the macroscopic synthetic equation. This offers substantial promise for the engineering applications of the DIG method.

\bibliographystyle{siamplain}
\bibliography{ref}

\end{document}